\documentclass{aa}

\usepackage{txfonts}
\usepackage{natbib}
\usepackage{graphicx}
\bibpunct{(}{)}{;}{a}{}{,} 

\begin{document}

   \title{The impact of \emph{Spitzer} infrared data on stellar mass
   estimates -- and a revised galaxy stellar mass function at
   $0 < z < 5$}
   
   \titlerunning{The Impact of \emph{Spitzer} infrared data on stellar
   mass estimates -- galaxy stellar mass function revised}
 
   \author{Franz~Elsner\inst{1,2},
           Georg~Feulner\inst{1,3,4}
	   \and
           Ulrich~Hopp\inst{1,3}
           }

   \authorrunning{Elsner et al.}

   \offprints{elsner@usm.lmu.de}

   \institute{Universit\"ats--Sternwarte M\"unchen, Scheinerstra\ss e
              1, D--81679 M\"unchen, Germany
              \and
              Max--Planck--Institut f\"ur
              Astrophysik, Karl--Schwarzschild--Stra\ss e 1,
              D--85748 Garching, Germany
              \and
              Max--Planck--Institut f\"ur
              extraterrestrische Physik, Giessenbachstra\ss e 1,
              D--85748 Garching, Germany
              \and
              Potsdam--Institut f\"ur Klimafolgenforschung, Postfach
              60~12~03, D--14412 Potsdam, Germany
	      }

   \date{Received 24 July 2007; accepted 29 October 2007}


  \abstract
   {}
   {We estimate stellar masses of galaxies in the high redshift
   universe with the intention of determining the influence of newly
   available Spitzer/IRAC infrared data on the analysis. Based on the
   results, we probe the mass assembly history of the universe.}
   {We use the \emph{GOODS--MUSIC} catalog, which provides multiband
   photometry from the U--filter to the 8~$\mu m$ Spitzer band for
   almost $15\,000$ galaxies with either spectroscopic (for $\approx
   7~\%$ of the sample) or photometric redshifts, and apply a standard
   model fitting technique to estimate stellar masses. We than repeat
   our calculations with fixed photometric redshifts excluding Spitzer
   photometry and directly compare the outcomes to look for systematic
   deviations. Finally we use our results to compute stellar mass
   functions and mass densities up to redshift $z = 5$.}
   {We find that stellar masses tend to be overestimated on average if
   further constraining Spitzer data are not included into the
   analysis. Whilst this trend is small up to intermediate redshifts
   $z \la 2.5$ and falls within the typical error in mass, the
   deviation increases strongly for higher redshifts and reaches a
   maximum of a factor of three at redshift $z \approx 3.5$. Thus, up
   to intermediate redshifts, results for stellar mass density are in
   good agreement with values taken from literature calculated without
   additional Spitzer photometry. At higher redshifts, however, we
   find a systematic trend towards lower mass densities if
   Spitzer/IRAC data are included.}
   {}


   \maketitle

\section{Introduction}

Whilst the assembly of stellar mass over cosmic history became a
matter of interest decades ago, substantial research on this problem
has become feasible only in the past few years. Extensive surveys had
to be carried out, and techniques to infer stellar masses of galaxies
from multicolor photometry had to be developed, as large complete
spectroscopic samples at cosmic distances are not yet available. In
general, these methods rest upon fitting a grid of stellar population
models to the data to compute stellar masses by multiplying the
mass--to--light ratio of the best matching template with its absolute
luminosity. This procedure has been used in numerous publications to
calculate stellar mass functions or stellar mass densities up to
intermediate redshifts \citep[e.g.][]{2000ApJ...536L..77B,
2001ApJ...562L.111D, 2003ApJS..149..289B, 2006A&A...453..869B}. With
the increasing availability of deeper surveys, the analysis has been
extended to higher redshifts, e.g. up to $z = 2$ or 3
\citep{2003ApJ...587...25D, 2003ApJ...594L...9F, 2003ApJ...599..847R,
2004A&A...424...23F}, or even to $z = 5$
\citep{2005ApJ...619L.131D}. However, the derived results are
typically based on observations from the U to the K--band and may be
affected by extrapolation errors as the rest--frame wavelength
coverage of the observed objects is shifted to the blue, and a good
observational constraint to a galaxy's optical to near--infrared
rest--frame luminosity is necessary to estimate its stellar mass
reliably. Results for stellar masses may therefore be impaired by
systematic uncertainties in the high redshift regime if observations
are not extended to longer wavelengths.

Since the Spitzer space telescope has become operational, it has been
possible to complement observations up to the K filter with
high--quality infrared data in the adjacent wavelength range
\citep{2004ApJS..154....1W}. To benefit from this improvement we used
the Great Observatories Origins Deep Survey, which provides deep
publicly available observations from numerous facilities in different
wavelength regimes \citep{dickinson}. Based on these data we probe the
influence of Spitzer photometry on the estimated stellar masses and
examine the consequences of the results on inferences about the mass
assembly history of the universe. In this work we address the effects
of Spitzer data to the mass calculation process only, i.e.\ we do not
study whether additional systematic deviations of photometric
redshifts emerge that, in general, can influence the result as
well. Furthermore we adopt a specific set of template models for the
calculations.

The paper is organized as follows. In Sect.~2 we give a short overview
of the dataset, and discuss the procedure adopted to estimate stellar
masses in Sect.~3. We focus on the influence of Spitzer data on the
derived masses in Sect.~4 and use the outcome to calculate stellar
mass functions (Sect.~5) and mass densities (Sect.~6). Finally, we
summarise our results and draw our conclusions in Sect.~7.

Throughout the paper we assume $\Omega_{\mathrm{M}} = 0.3$,
$\Omega_{\Lambda} = 0.7$ and $H_{0} =
70~\mathrm{km\,s^{-1}\,Mpc^{-1}}$. Magnitudes are given in the AB
system.

\section{The dataset}

To study the assembly of stellar mass we focus on the Chandra Deep
Field South, where numerous observations within the framework of the
Great Observatories Origins Deep Survey (\emph{GOODS}) provide data
over a wide range of wavelengths. The present paper is based on
\emph{GOODS--MUSIC}, a multicolor catalog published by
\citet{2006A&A...449..951G}. We briefly discuss its main
characteristics here and refer the reader to that paper for a more
detailed description.

The catalog combines U--band data from the 2.2m--MPE/ESO telescope and
VLT/VIMOS ($\mathrm{U_{35}}$, $\mathrm{U_{38}}$ and
$\mathrm{U_{VIM}}$), Hubble/ACS images in F435W (B), F606W (V), F775W
(i) and F850LP (z), VLT/ISAAC data in J, H and $\mathrm{K_S}$, and
Spitzer/IRAC data at $3.6~\mu m$, $4.5~\mu m$, $5.8~\mu m$ and
$8.0~\mu m$. Since observations have not yet been finished in all
bands, the coverage fraction lies at around 63~\% for
$\mathrm{U_{VIM}}$--data and at about 54~\% for the H--band (ISAAC
data release 1.0). Source detection has been performed independently
in both the deep z--image (14651 objects detected) and the shallower
$\mathrm{K_S}$--band data (2931 sources), ending up with a z-- and
$\mathrm{K_S}$--complete catalog consisting of $14\,847$ objects in
total. Special software was developed for photometry
\citep{2006MSAIS...9..454D} to meet the requirements of color
measurement in combined ground and space based observations with
dissimilar point spread functions. Out of all objects, 13767, 12041,
6767 and 5869 sources could be detected in the IRAC data in channels
1, 2, 3 and 4, respectively. For objects undetected in a specific
image, upper limits in flux were calculated on the basis of
morphological information derived from the detection image. As a last
step redshift information were added to complete the dataset. To do
this, spectroscopic surveys available at that time were used to assign
1068 spectroscopic redshifts. For the remaining objects, a standard
photometric redshift code was applied that was able to reproduce the
spectroscopic redshifts with an accuracy of $\langle|\frac{\Delta z}{1
+ z}|\rangle = 0.045$.  As shown in Fig.~12 of
\citet{2006A&A...449..951G}, less than 2~\% of the spectroscopically
observed objects reveal a photometric redshift that deviates severely
($|z_{spec} - z_{phot}| > 0.3 = 5 \cdot \langle \sigma \rangle$) from
the spectroscopic one.

In summary, the catalog used here consists of $14\,847$ objects
enclosing at least 72 stars and 68 AGNs over a total area of
143.2~$\mathrm{arcmin^2}$, with mean limiting magnitudes of
$\mathrm{z_{lim}} \approx 26.0$ and $\mathrm{K_{S \; lim}} \approx
23.8$ at 90~\% completeness level.

\section{Deriving stellar masses}

To calculate galaxy stellar masses we adopted the method described in
\citet{2004ApJ...608..742D} and locally tested against spectroscopic
results in \citet{2004ApJ...616L.103D}. This method is based on
comparing object colors to those of a template library of stellar
population synthesis models. The five-dimensional model grid used here
was computed from synthetic \citet{bruzual-2003-344} models with an
underlying Salpeter initial mass function (IMF) truncated at 0.1 and
100~M$_{\sun}$. It is parameterized by a star formation history (SFH)
of the form $\psi(t) \propto \mathrm{exp}(-t/\tau)$ evaluated at
$\tau$ = \{0.5, 1.0, 2.0, 3.0, 5.0, 8.0, 20.0\}~Gyr at 15 different
ages $t \in [0.2, 10.0]$~Gyr, with dust extinction between
$A_\mathrm{V} = 0$ and $A_\mathrm{V} = 1.5$ magnitudes using a
\citet{2000ApJ...533..682C} extinction law. While covering the
physical relevant parameter range in $t$ and $\tau$, the upper limit
in $A_\mathrm{V}$ must be considered as a restriction. We adopted this
to take into account the degeneracy in age and extinction to suppress
solutions with unexpectedly high values for $A_\mathrm{V}$. However,
our sample may contain a small number of heavily dust enshrouded
galaxies that, in turn, will not be treated appropriately. In addition
to the main component, a starburst was superimposed which was allowed
to contribute at most 20~\% to the z--band luminosity in
rest--frame. It was modeled as a 50~Myr old episode of constant star
formation with an independent extinction up to $A_\mathrm{V} (b) =
2.0$ magnitudes. To take into account polycyclic aromatic hydrocarbon
(PAH) emission, which becomes important at rest--frame wavelength
$\lambda \ga 6~\mu m$ and is attributed to starforming regions, the
spectral energy distribution (SED) of the burst component was modified
by including PAH--emission features following
\citet{2005ApJ...619..755D}. We found that the object fluxes are
reconstructed slightly better at low redshifts compared to a
burst--SED lacking this feature, while the actual result for
calculated stellar masses is not significantly affected ($|\log \,
M_\mathrm{with \; PAH}/M_\mathrm{without \; PAH}| <
0.003~\mathrm{dex}$ at $z < 1$). Because of the well known
age--metallicity degeneracy, we restricted our models to solar
metallicities and performed tests to ensure that we do not introduce
significant systematic deviations with this constraint.

\begin{figure}
  \resizebox{\hsize}{!}{\includegraphics{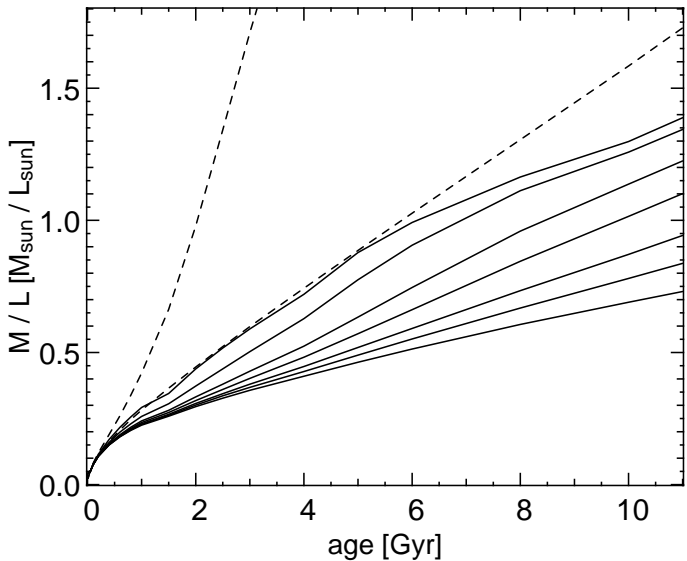}}
  \caption{Mass--to--light ratio of our template models in solar
  units. The evolution of $M/L_\mathrm{K}$ for different SFH $\tau$ =
  \{0.5, 1.0, 2.0, 3.0, 5.0, 8.0, 20.0\}~Gyr (\emph{solid lines, from
  top to bottom}) is shown. For comparison, the much wider dynamical
  range of the corresponding $M/L_\mathrm{V}$ ratio is also indicated
  (\emph{area between dashed lines}).}
  \label{fig:ml}
\end{figure}

To derive stellar masses, we used photometry in the filters
$\mathrm{U_{38}}$, B, V, i, z, J, H, $\mathrm{K_S}$ and the four IRAC
channels. In the blue we focused on $\mathrm{U_{38}}$ only, because
the $\mathrm{U_{35}}$--filter is known to be leaking and the
$\mathrm{U_{VIM}}$ observations do not cover the whole field
\citep[see][]{2006A&A...449..951G}. We want to emphasize here that the
Spitzer observations provide an unprecedented opportunity to include
high--quality infrared photometry longward of about $3~\mu m$ in
stellar mass estimates. For each object we computed the full
likelihood distribution of our models shifted to the corresponding
redshift. To infer the most probable mass--to--light ratio
($M/L_\mathrm{K}$) we weighted the individual $M/L_\mathrm{K}$ ratios
of our templates by their likelihoods and averaged over all parameter
combinations. Moreover, we were able to derive an estimate of the
expected error of this quantity from the width of the distribution. By
utilizing the $\mathrm{K_S}$--band $M/L$ ratio we benefit from several
advantages. In general, the variation with age in $M/L_\mathrm{K}$ is
small compared to its optical counterpart as shown in
Fig.~\ref{fig:ml}. Furthermore, dust absorption is small at longer
wavelengths, so potentially large uncertainties in $A_\mathrm{V}$ do
not alter the result substantially. The actual mass, which was
calculated by multiplying the $M/L_\mathrm{K}$ ratio with the
k--corrected total $\mathrm{K_S}$--band luminosity of the best fitting
SED model, turns out to be comparatively robust with a mean error of
$\sigma_\mathrm{log \, M} = 0.16 \; dex$. Besides the error attributed
to the template fitting process only, calculated stellar masses are
affected by further important sources of uncertainty. To take them
into account we performed extensive Monte Carlo simulations. We
analyzed 1000 simulated catalogs where we have considered errors in
the object redshifts, calculated $M/L_\mathrm{K}$--ratios and a
general uncertainty in photometry. The errors were propagated to the
results for stellar masses and the mean uncertainty was estimated to
be about $\sigma_\mathrm{log \, M} = 0.33 \; dex$.

\begin{figure}
  \resizebox{\hsize}{!}{\includegraphics{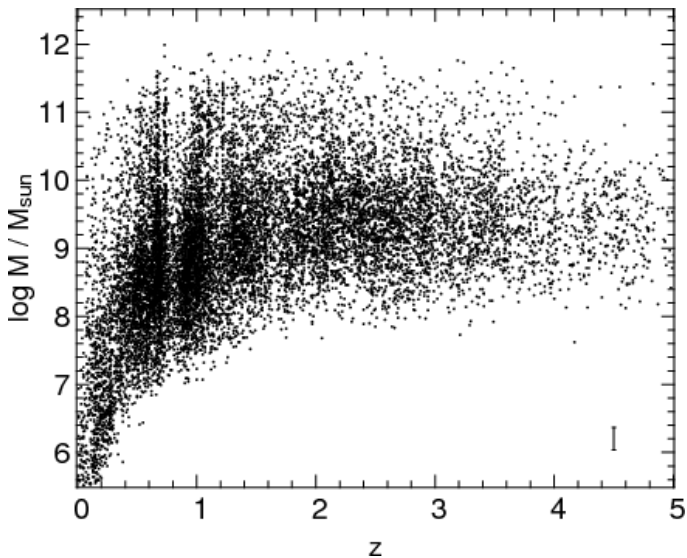}}
  \caption{Stellar masses as a function of redshift. The typical error
  in mass is indicated in the lower right-hand corner.}
  \label{fig:mz}
\end{figure}

In Fig.~\ref{fig:mz} we show the distribution of all galaxies in the
mass versus redshift plane. We found five galaxies at high redshifts
$z > 4$ with masses $\log\,M/M_{\sun} > 11.5$. The presence of such
high--mass objects less than 2~Gyr after the big bang is still
controversial \citep[see, for example,][and references
therein]{2006astro.ph..6192D}. However, a closer analysis reveals that
they are more likely to be heavily dust enshrouded galaxies at
intermediate redshifts. While keeping the original upper limit in
$A_\mathrm{V}$ for the remaining sample, to account for the
degeneracy in age and extinction as explained above, we extended the
parameter space by allowing a larger maximal dust extinction of
$A_\mathrm{V} = 4$ magnitudes for these five galaxies. Refitting the
objects' redshifts with our full template library leads to much
smaller redshifts $z \approx 2$ and typical masses about
$\log\,M/M_{\sun} \approx 10.5$ for four of the galaxies. With this
reanalysis, the fit quality of our best matching SED template
increases significantly and becomes comparable to the mean for these
four objects. The fifth galaxy remains at high $z$ primarily because
of an unexpected low $\mathrm{K_S}$--band luminosity, which may be a
result of incorrect photometry. Nevertheless, we want to emphasize
that it is not possible to draw an unambiguous conclusion about the
nature of these objects on the basis of currently available data. This
uncertainty reflects the general character of samples with mainly
photometric redshifts. While they can provide accurate data within a
statistical analysis, for a dedicated study of rare object properties
additional information is still required. To restrict our sample to a
uniform parameter space, we excluded the five questionable galaxies
from our analysis in Sect.~4.

\section{The role of \emph{Spitzer}}

To study the impact of Spitzer/IRAC data on our analysis, we
considered only those galaxies with errors in mass from the fitting
process $\sigma_\mathrm{log \, M} < 0.2$ ($\approx 13\,000$). With
this restriction, about half of the objects were detected in at least
three IRAC channels. However, even if a source has not been identified
in a specific filter, an upper limit in flux was included into the
analysis, which provides a valuable additional constraint and in
general affects the derived stellar mass. For this sample we repeated
the calculation of stellar masses as described in Sect.~3 but
reduced photometric information to the filters $\mathrm{U_{38}}$, B,
V, i, z, J, H and $\mathrm{K_S}$ only, i.e.\ excluded the four IRAC
channels. This subset of our data represents a multi--band catalog
with wavelength coverage from about 3500~\AA\ to $23\,000$~\AA,
which was typically available to study the high redshift universe
before facilities such as the Spitzer space telescope became
operational. We were then able to study the influence of additional
infrared photometry by comparing the properties of the best--fitting
models directly.

\begin{figure}
  \resizebox{\hsize}{!}{\includegraphics{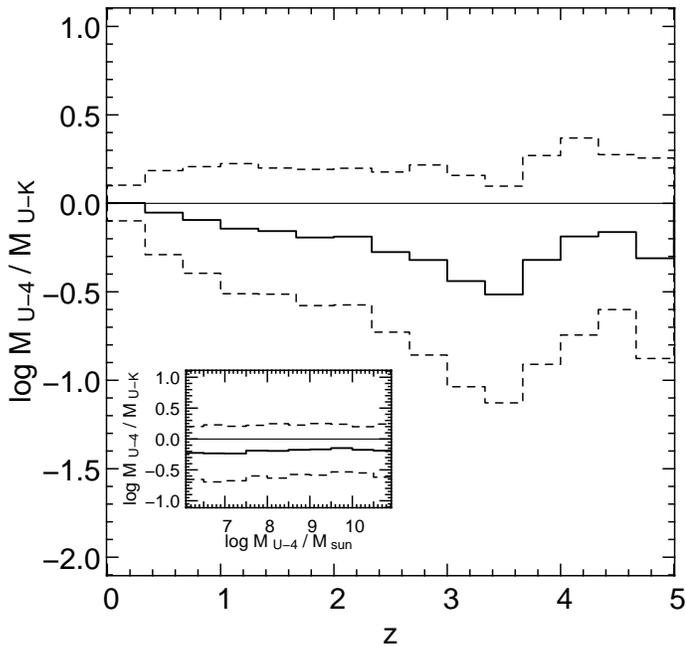}} \caption{This
  figure shows the difference in masses calculated under the inclusion
  of the Spitzer bands ($M_\mathrm{U-4}$) and with restricted
  photometry ($M_\mathrm{U-K}$) as a function of redshift.  Averaged
  over all redshifts, the dependency on stellar mass itself is also
  indicated (\emph{inset}). Mean values (\emph{solid lines}) and
  r.m.s. of the data--points (\emph{dashed lines}) are shown.}
  \label{fig:dm}
\end{figure}

We found stellar masses to be overestimated without Spitzer data on
average. The mean deviation between the masses calculated with all
filters up to IRAC channel~4 ($M_\mathrm{U-4}$) and those with limited
photometry ($M_\mathrm{U-K}$) of the whole sample is $\log \,
M_\mathrm{U-4}/M_\mathrm{U-K} = -0.18~\mathrm{dex}$ with a r.m.s. of
0.41. As shown in Fig.~\ref{fig:dm}, this trend is relatively
independent of mass itself, but has a strong dependency on
redshift. Whilst the mean deviation up to a redshift of $z \approx 2$
is only moderate ($|\log \, M_\mathrm{U-4}/M_\mathrm{U-K}| <
0.2~\mathrm{dex}$) and comparable to the typical error in mass, the
difference reaches a maximum of $|\log \,
M_\mathrm{U-4}/M_\mathrm{U-K}| \approx 0.5~\mathrm{dex}$ at $z \approx
3.5$. This corresponds to an overestimation of mass of more than a
factor of three on average, whereas the scatter for individual sources
is large in general, as indicated by the r.m.s. of the
distribution. Therefore we can confirm the conjecture made by
\citet{2006A&A...459..745F}, who did the same analysis on the
$\mathrm{K_S}$--selected subsample with about 3000 galaxies, although
we found the effect of Spitzer/IRAC data to be somewhat more
pronounced. At higher redshifts the difference in mass decreases
again. We can therefore conclude directly that without further data
points the extrapolation of object luminosities to longer wavelengths
is not straight forward and induces systematic deviations. This result
has been derived for a specific set of template SEDs with solar
metallicity, a Calzetti extinction law, Salpeter IMF, and for fixed
photometric redshifts. Although other choices of model parameters may
result in different absolute values for stellar masses, we expect the
detection of a systematic trend to be comparatively robust, as it
depends on a relative deviation. This statement is supported by the
results of \citet{2006A&A...459..745F} who found a similar trend
despite using a different model template library and SED fitting
procedure.

\begin{figure}
  \resizebox{\hsize}{!}{\includegraphics{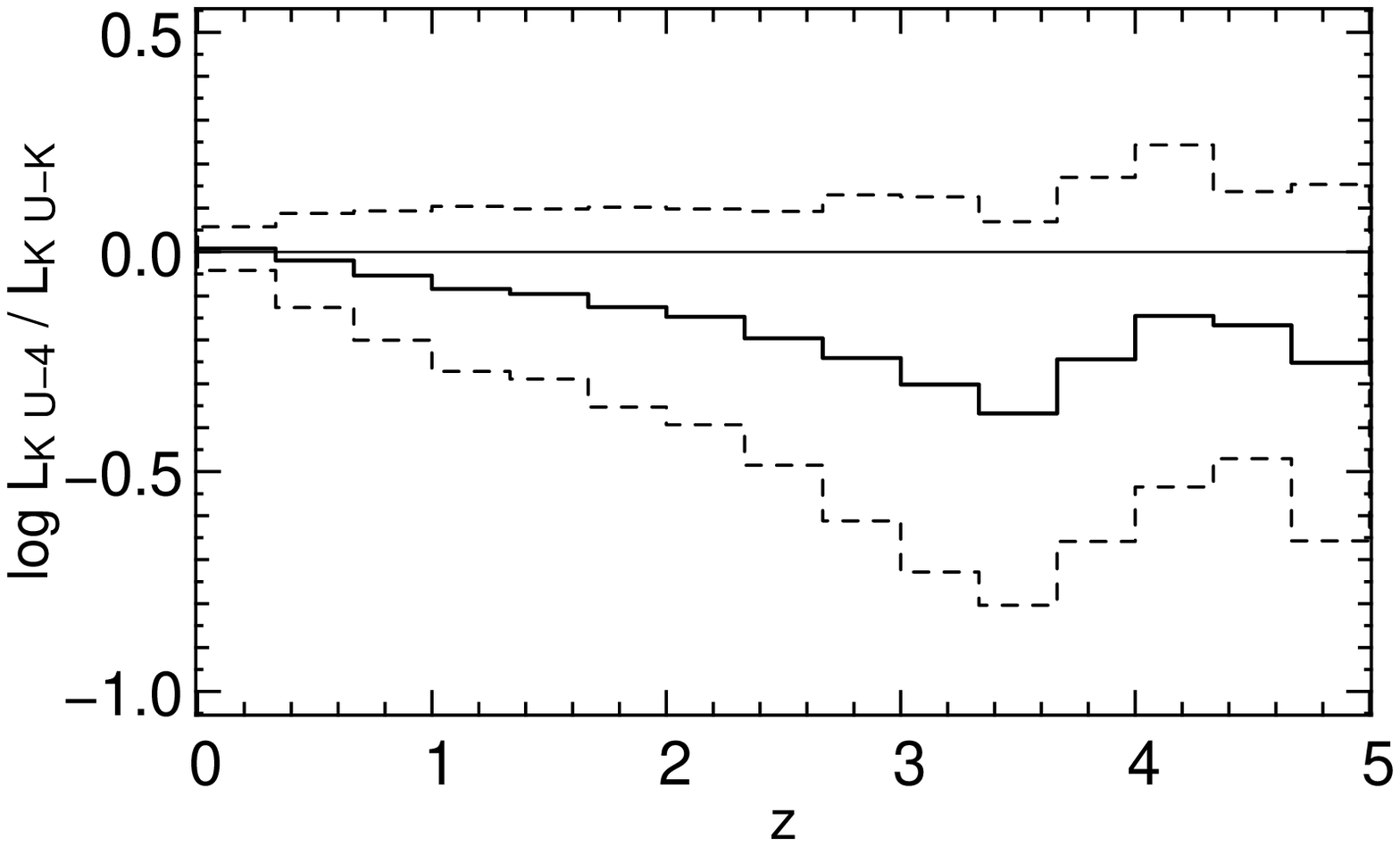}}
  \resizebox{\hsize}{!}{\includegraphics{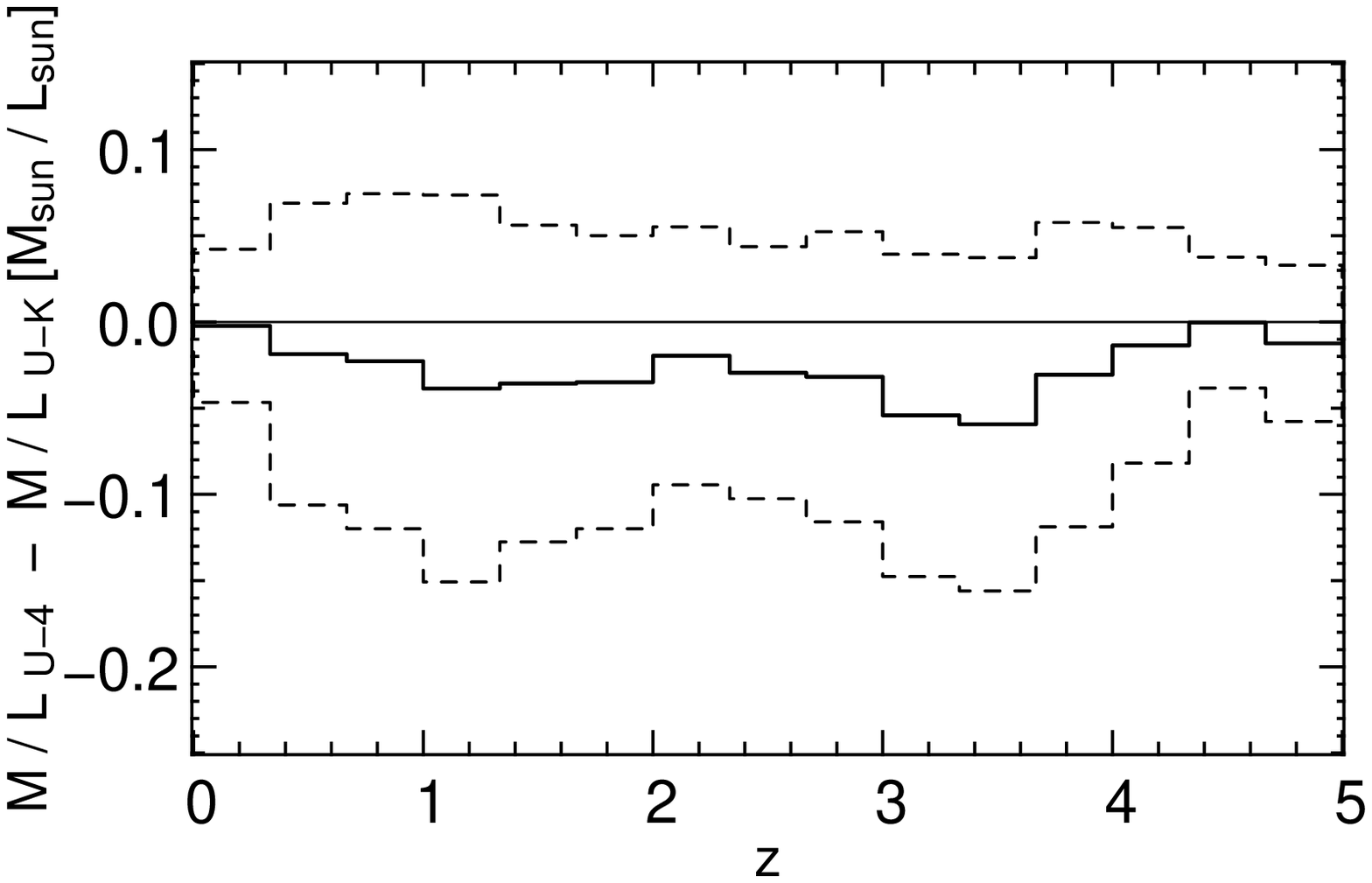}}
  \caption{Ratio of inferred absolute $\mathrm{K_S}$--band
  luminosities (\emph{upper panel}) and shift in corresponding
  mass--to--light ratios in solar units (\emph{lower panel}) as a
  function of redshift. The mean values (\emph{solid lines}) and the
  r.m.s. of the data--points (\emph{dashed lines}) are shown.}
  \label{fig:dmzul}
\end{figure}

\begin{figure}
  \resizebox{\hsize}{!}{\includegraphics{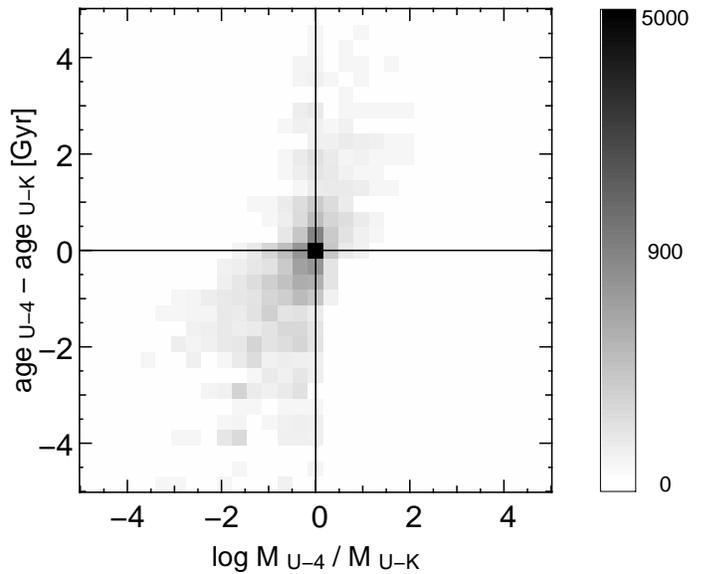}}
  \caption{Correlation between the shift in masses (\emph{x--axis})
  and that in ages (\emph{y--axis}) after inclusion of Spitzer
  photometry. The frequency of a specific combination is indicated by
  the color in nonlinear scaling. The correlation coefficient of the
  two quantities is $r = 0.6$.}
  \label{fig:korr}
\end{figure}

The immediate reasons for the deviation can be seen in
Fig.~\ref{fig:dmzul}, where shifts in both the underlying
$\mathrm{K_S}$--band luminosities and the mass--to--light ratios are
evident. For a robust calculation of stellar masses with the aid of
infrared luminosities, a good knowledge of the slope of the spectra
beyond the $4000$~\AA\ break is mandatory. Whereas the brightness of
an old massive galaxy decreases slowly in this wavelength range, a
young low--mass object has a steep spectrum and the decline is much
stronger. To distinguish between early and late type galaxies solely
on the basis of the blue part of their spectra is difficult, because
the possible presence of a starburst can alter a galaxy's UV
luminosity substantially. Furthermore, uncertainties in extinction
strongly affect this wavelength range, which makes it even more
difficult to reveal the fundamental properties of the underlying
SED. Therefore several data points longward of $4000$~\AA\ are
required for a reliable estimation of stellar masses. However, at a
redshift of $z = 2.5$ the 4000~\AA\ break lies at $14\,000$~\AA ,
i.e.\ it has already passed the J filter, so the calculation is
resting upon H and $\mathrm{K_S}$--band photometry predominantly, as
the slope of the rest--frame optical/near--infrared part of a galaxy's
spectrum is characterized by the $\mathrm{H} - \mathrm{K_S}$ color
only. At a redshift of $z \approx 3.5$, where the deviation is
largest, the spectrum longward of the $4000$~\AA\ break in rest--frame
is covered solely by the $\mathrm{K_S}$--filter and is therefore
insufficiently sampled. In general, photometry becomes increasingly
uncertain at higher redshifts when objects fade, so the constraints on
the slope weaken further. At this redshift, the systematic deviation
in masses due to the inclusion of Spitzer data starts to dominate over
possible intrinsic errors. This effect can be traced back to the
assignment of model SEDs that are too old with too high an absolute
infrared luminosity. A decrease in stellar mass is therefore often
associated with attributing a younger model SED, i.e.\ there is a
correlation between shift in masses and ages, as can be seen in
Fig.~\ref{fig:korr}. Since extinction influences the shape of the
spectrum in a similar way to age, the degeneracy in the two quantities
reduces the correlation. We show the change of the best fitting model
SED for several galaxies in the appendix.

The reasons for decreasing differences in stellar masses for even
higher redshifts beyond $z = 3.5$ are twofold. First, the maximal age
of a possible fit model is restricted by the age of the universe at
that redshift, which acts as an upper limit. Therefore, the fit models
are forced to be younger and a large decline in age with a lower
resulting mass is therefore not likely. Secondly, at these extreme
redshifts sources are very faint and an increasing fraction of our
sample is hardly detected in the shallower J, H, and
$\mathrm{K_S}$--band data. In this case, the calculation of masses
without Spitzer data is mainly based on the V, i and z photometry (as
galaxies are U and B--filter dropouts) and it turns out that very
young low--mass model SEDs were assigned. With the inclusion of
Spitzer bands the masses for this type of objects tend to increase,
and the average difference of the whole sample gets smaller. We want
to mention here that whereas the former effect is universal, the
latter is a property of our specific catalog and may be less
pronounced in other surveys.

\begin{figure}
  \resizebox{\hsize}{!}{\includegraphics{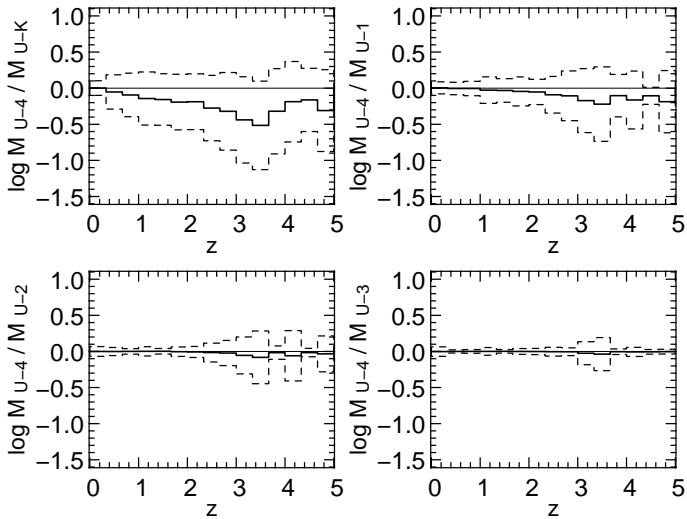}}
  \caption{Deviation in mass as a function of redshift after including
    Spitzer filters one by one. Starting from masses calculated
    without Spitzer photometry (\emph{upper left panel, same as
    Fig.~\ref{fig:dm}}), the remaining deviation is shown for
    calculations based on photometry including Spitzer/IRAC data
    (\emph{up to channel 1, upper right; to channel 2, lower left; to
    channel 3, lower right}). Mean values (\emph{solid lines}) and the
    r.m.s. of the data--points (\emph{dashed lines}) are plotted.}
    \label{fig:stin}
\end{figure}

In the next step we examined the influence of single Spitzer filters
on the resulting stellar mass estimates. We repeated our calculations
after adding the four Spitzer/IRAC bands to the analysis one by
one. The outcome is illustrated in Fig.~\ref{fig:stin} where we show
deviations in mass in relation to those calculated with full
photometric information. Only one further data point at $3.6~\mu m$
can reduce the remaining shift in masses to less than $|\log \,
M_\mathrm{U-4}/M_\mathrm{U-1}| \le 0.25~\mathrm{dex}$ over the full
redshift--range. After additionally taking into account Spitzer
channel 2 at $4.5~\mu m$, the systematic deviation shrinks to $|\log
\, M_\mathrm{U-4}/M_\mathrm{U-2}| \le 0.1~\mathrm{dex}$ and thus
becomes comparable to possible intrinsic errors attributed to the
method used to derive stellar masses. Similarly, it is possible to
reduce the deviation including just Spitzer channels 3 and 4 in the
calculation. Although the remaining difference is at most $|\log \,
M_\mathrm{U-4}/M_\mathrm{U-K+3-4}| \le 0.24~\mathrm{dex}$ and
therefore larger than in the case discussed above, and taking into
account lower quality data, providing upper limits in flux can often
significantly improve the result. Despite reduced mean deviations, the
change in mass for single objects can still be large.

\section{The stellar mass function}

Because we found systematic shifts in calculated stellar masses, we
wanted to examine the effects on resulting quantities such as the
stellar mass function, i.e.\ the number of objects per comoving volume
and mass interval. To do so, we subdivided our catalog into seven
redshift bins from $z = 0.25$ to $z = 5$. Using this relatively coarse
grid we still work with 380 objects in the highest $z$--bin and ensure
that our results are based on statistically meaningful samples. Within
each interval we calculated stellar mass functions after correcting
our data points for incompleteness due to the flux--limited object
selection utilising the $V/V_\mathrm{max}$--formalism of
\citet{1968ApJ...151..393S}. In this context it is important to point
out that we do not achieve a well defined completeness limit in
stellar mass, as even for a sharp underlying flux limit, the dynamic
range of mass--to--light--ratios results in a wide distribution in
mass. Therefore the largest $M/L$--ratio at a certain redshift
determines the limit in mass below which incompleteness starts to play
a role. To quantify this effect we followed two different
approaches. First, motivated by a more theoretical point of view, we
studied the evolution of a passive evolving galaxy with subsolar
metallicity and zero extinction formed at a redshift of $z = 10$ in a
single burst. By assuming an underlying flux limit of
$\mathrm{z_{lim}} = 26.0$~magnitudes\footnote{The magnitude limits
were calculated in \citet{2006A&A...449..951G} using simulations. In
the z--band, the limit varies little over the field as the exposure
map is relatively homogeneous. The variation in the
$\mathrm{K_S}$--filter is larger.}  we computed the corresponding
stellar mass as a more conservative estimate for the completeness
limit of the catalog. We found that galaxies with masses down to
$M\mathrm{_{lim}} \approx 3 \cdot 10^9~M_{\sun}$ can be detected at $z
= 1$. The value is increasing to $M\mathrm{_{lim}} \approx 3 \cdot
10^{10}~M_{\sun}$ at $z = 2.5$ where those objects would be detected
in the shallower $\mathrm{K_S}$--band with an average flux limit of
$\mathrm{K_{S \; lim}} = 23.8$~magnitudes. This limit in mass is also
obtained for old but moderately star--forming systems with extinctions
around $A_V \approx 1$. In the second approach, which is motivated by
the data actually available, we studied the evolution of the
mass--to--light ratios of the catalog and calculated the 95~\%
quantile in $M/L_z$ as a function of redshift e.g. the limit below
which 95~\% of the $M/L_z$ ratios of our sample are located. Assuming
a sharp flux limit such as adopted above, we computed the
corresponding stellar mass as an estimate for completeness. For a
redshift of $z = 1$ we found this value to be $M\mathrm{_{lim}}
\approx 2 \cdot 10^9~M_{\sun}$ increasing to $M\mathrm{_{lim}} \approx
9 \cdot 10^9~M_{\sun}$ at $z = 2.5$. Although this more aggressive
method results in a lower mass limit, it should be clear that a
significant fraction of heavily dust enshrouded galaxies with large
$M/L$--ratios may stay undetected at intermediate $z$, while a
reliable detection of typical more massive active galaxies with low to
intermediate extinctions should be possible up to high redshifts. As
the two methods result in different completeness limits we performed
the calculations in this section for both values independently.

\begin{figure*}
  \includegraphics[width=17cm]{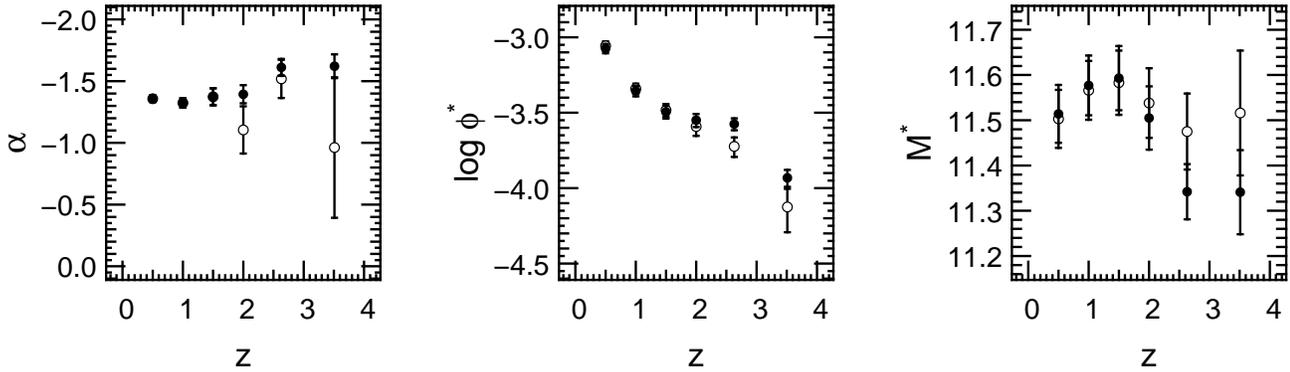}
  \caption{Evolution of Schechter parameters of the stellar mass
  function with redshift, showing the slope--parameter (\emph{left
  panel}) as derived from an independent fit in each redshift
  interval. The remaining parameters $\phi^*$ (\emph{center}) and
  $M^*$ (\emph{right panel}) were recalculated using a fixed value for
  $\alpha$. They are interdependent in that a shift in $M^*$ to higher
  values, for example, would result in a smaller value of
  $\phi^*$. The results using mass limits from a $M/L$--ratio analysis
  (\emph{filled circles}) and a passively evolving scenario
  (\emph{open circles, see text}) are plotted.}
  \label{fig:spar}
\end{figure*}

To estimate the errors of the stellar mass function we used 1000
realisations of randomly drawn catalogs for which we have considered
errors in computed redshifts, calculated $M/L_\mathrm{K}$--ratios and
a general uncertainty in photometry. Finally, to cope with
incompleteness at lower masses, in view of our subsequent analysis, we
fitted our datapoints from the massive end down to the completeness
limit with an analytical expression suggested by
\citet{1976ApJ...203..297S} of the form:
\begin{displaymath}
   \psi(M; \phi^*, M^*, \alpha) = log(10) \cdot \phi^* \cdot \left[ 10^{(M - M^*)} \right] ^{(1+\alpha)} \cdot exp\left[-10^{(M - M^*)}\right]
\end{displaymath}
In this formula the number density $\psi(M)$ is parameterized via a
scale factor $\phi^*$, a typical mass $10^{M^*}M_{\sun}$ and a
slope--parameter $\alpha$. First we computed the values of the fit
parameters through an $\chi^2$--analysis in the redshift bins up to $z
= 4$ independently. We excluded the highest redshift interval at $4
\le z < 5.01$ from the procedure, as we found the Schechter function
to be insufficiently constrained due to the increasing mass
limit. However, a robust estimation of the parameters is difficult as
they are highly degenerated in the expression used. To deal with this
problem we decided to fix the slope--parameter at its error--weighted
mean value, as we found no clear evidence for an evolution with
redshift as shown in the left--hand panel of Fig.~\ref{fig:spar}. We
also adopted this procedure because an undersampling of low mass
objects in the catalog may affect the determination of the slope at
higher redshifts in a systematic manner. With this constant value of
$\alpha$ we repeated our calculation of the two remaining parameters
$M^*$ and $\phi^*$ in every $z$--bin; for both completeness limits the
result is plotted in Fig.~\ref{fig:spar} and listed in
Table~\ref{tab:spar}. A relatively uniform decrease in $\phi^*$ with
redshift is clearly evident, which reflects the fact of a general
decline in the number of detected objects in place. In contrast to the
evident decrease of $\phi^*$, it is hard to say whether there is a
hint at a mass dependent evolution of the number density, which would
manifest itself as a shift in the parameter $M^*$. Although this trend
would be expected by the downsizing scenario for galaxy evolution,
where massive galaxies tend to form earlier than their low--mass
counterparts, a slight increase in the typical mass with redshift
followed by a decrease as indicated in the plot may also be the result
of large--scale structure within the observed field.

\begin{figure*}
  \includegraphics[width=17cm]{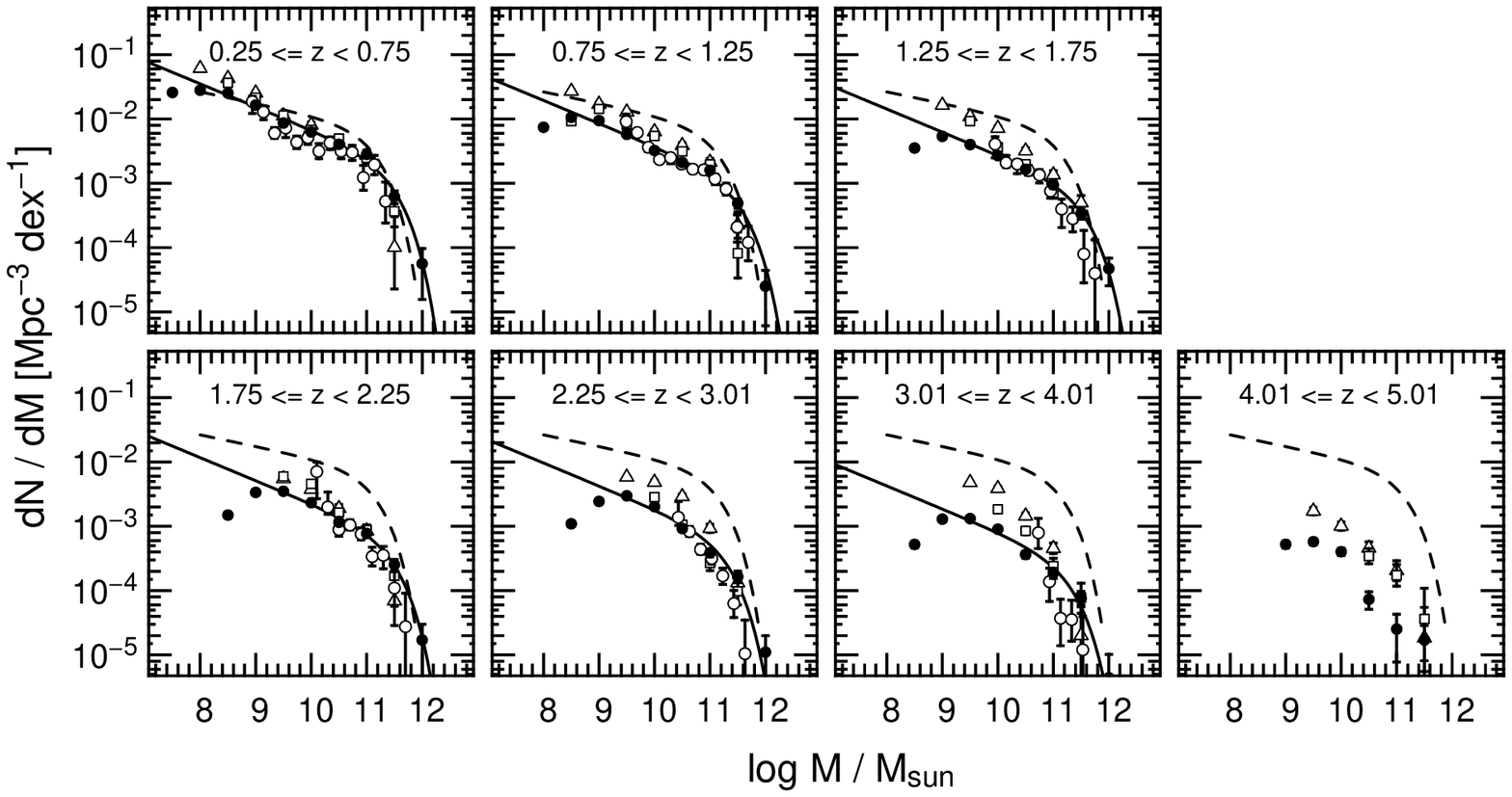}
  \caption{Evolution of stellar mass function with redshift, showing
  data points computed using the $V/V_\mathrm{max}$--formalism in
  seven different redshift intervals (\emph{filled circles}) and, for
  $z \le 4$, the best fitting Schechter functions utilizing mass
  limits from a $M/L$--ratio analysis (\emph{solid line, see
  text}). For comparison we also plot the result of
  \citealt{2005ApJ...619L.131D} (\emph{open triangles FDF, open
  squares GSD}) derived without additional Spitzer infrared data and
  the result of \citet{2006A&A...459..745F}, calculated within
  slightly different redshift bins up to $z \le 4$ (\emph{open
  circles}), and the local Schechter function of
  \citet{2001MNRAS.326..255C} (\emph{dashed line}).}
  \label{fig:smf}
\end{figure*}

The Schechter functions can be seen in Fig.~\ref{fig:smf}, where we
show the fit to the data with respect to the local stellar mass
function of \citet{2001MNRAS.326..255C} with the parameters $M^*$ =
11.16, $\phi^*$ = 0.0031 and $\alpha$ = -1.18. For comparison we also
plot the mass function of \citet{2006A&A...459..745F}, as derived from
the $\mathrm{K_S}$--selected subsample of the \emph{GOODS--MUSIC}
catalog in slightly varied redshift bins up to $z \le 4$. Besides a
tendency to a smaller number of high mass objects, the datasets show a
general agreement. The discrepancy at the high mass end turns out to
be robust. Although results in this region rely on only a few
galaxies, the error in mass is small for most of them. We discuss
possible reasons for the deviations in detail in the next section. We
also display the result of \citet{2005ApJ...619L.131D} calculated from
the \emph{FORS Deep Field (FDF)} and a subarea of the \emph{GOODS--S}
(\emph{hereafter GSD}) in the same redshift intervals without
additional Spitzer infrared data. While the \emph{GSD} data can be
reproduced well up to intermediate redshifts, the analysis of the
\emph{FDF} sample tends to result in larger values for stellar mass
function. At high redshifts, $z > 3$, deviations are clearly visible
for both the \emph{FDF} and \emph{GSD} datasets. They can be
attributed to distinct effects; in addition to a shift in the mass
scale as a result of the influence of Spitzer photometry, the total
number of objects in the highest redshift intervals derived from an
integral over the mass function is larger. Performing a
Kolmogorov--Smirnov test on the redshift distributions of the
\emph{FDF} galaxies and the \emph{GOODS--MUSIC} catalog used in this
work, we can clearly reject the hypothesis that the two samples are
drawn from the identical parent distribution at a 1~\% level. However,
both effects can have the same origin since the calculation of
photometric redshifts becomes less reliable at high redshifts if the
spectra are insufficiently constrained in the rest--frame optical,
especially as spectroscopic redshifts used for comparison become very
rare and are restricted to the most luminous objects.

\begin{table*}
   \centering
   \caption{Schechter parameters of stellar mass functions
   as derived using the mass limits from a $M/L$--ratio analysis
   (\emph{upper section}) and a passively evolving scenario
   (\emph{lower section, see text}).}
   \label{tab:spar}
   \begin{tabular}{ccccccc}
      \hline
      \hline
      redshift interval & $M^*$ & $\sigma_{M^*}$ & $\phi^* \cdot 10^4$ & $\sigma_{\phi^*} \cdot 10^4$ & $\alpha$ & $\sigma_{\alpha}$\\
      \hline
      $0.25 \le z < 0.75$ & 11.51 & 0.06 & 8.43 & 0.60 & -1.358 & 0.023\\
      $0.75 \le z < 1.25$ & 11.58 & 0.07 & 4.41 & 0.35 & -1.358 & 0.023\\
      $1.25 \le z < 1.75$ & 11.59 & 0.07 & 3.20 & 0.31 & -1.358 & 0.023\\
      $1.75 \le z < 2.25$ & 11.51 & 0.07 & 2.82 & 0.28 & -1.358 & 0.023\\
      $2.25 \le z < 3.01$ & 11.34 & 0.06 & 2.66 & 0.24 & -1.358 & 0.023\\
      $3.01 \le z < 4.01$ & 11.34 & 0.10 & 1.17 & 0.15 & -1.358 & 0.023\\
      \hline
      $0.25 \le z < 0.75$ & 11.50 & 0.06 & 8.77 & 0.62 & -1.352 & 0.023\\
      $0.75 \le z < 1.25$ & 11.57 & 0.07 & 4.57 & 0.36 & -1.352 & 0.023\\
      $1.25 \le z < 1.75$ & 11.58 & 0.07 & 3.29 & 0.32 & -1.352 & 0.023\\
      $1.75 \le z < 2.25$ & 11.54 & 0.08 & 2.56 & 0.34 & -1.352 & 0.023\\
      $2.25 \le z < 3.01$ & 11.48 & 0.08 & 1.89 & 0.28 & -1.352 & 0.023\\
      $3.01 \le z < 4.01$ & 11.51 & 0.14 & 0.75 & 0.24 & -1.352 & 0.023\\
      \hline
      \hline
   \end{tabular}
\end{table*}

\section{The stellar mass density}

We are now able to compute stellar mass densities, i.e.\ the mass in
stars and remnants per comoving volume at a specific redshift, on the
basis of our results from Sect.~5. For this calculation we divided our
stellar mass functions in each redshift bin into two mass intervals at
the threshold values of the two completeness limits considered
here. Above the limit, where data points and Schechter function fall
together, we summed the stellar masses of our objects directly. In
contrast to this procedure, we integrated the Schechter function in
the low--mass range down to zero to take account of the fact that our
catalog suffers from incompleteness here. The completion to lower
masses contributes about 13~\% (43~\%) to the final mass density at
the redshift interval $3.01 \le z < 4.01$ using the mass limits
derived from a $M/L$--ratio analysis (passive evolution scenario). In
the highest redshift bin we summed stellar masses directly, not
correcting for completeness. Although we certainly underestimate the
resulting outcome for stellar mass density, a comparison to the values
published by \citet{2005ApJ...619L.131D} is still possible as the
results were calculated without corrections there. In order to check
our results for robustness, we dropped our assumption of a constant
slope parameter $\alpha$ and recalculated stellar mass densities, but
did not find appreciable deviations. To assign errors to the resulting
data points we again used Monte Carlo simulations, and additionally
considered uncertainties in the Schechter--parameters that affect the
contribution of the integral over the low mass range only. However, a
more careful analysis reveals that resulting errors may not include
all sources of uncertainty. For example, cosmic variance can alter our
result on a 20~\% level when estimating the expected uncertainty in
number density of observed objects \citep{2004ApJ...600..171S},
although we were able to draw on a relatively large survey area of
about 140~$\mathrm{arcmin^2}$ for our calculations. Another source of
uncertainty is the proper treatment of stars in the post--AGB phase
and their influence on stellar population synthesis models \citep[see,
for example,][]{2006ApJ...652...85M, 2007astro.ph..2091B,
2006ApJ...652...97V}. Furthermore, deviations from the assumed IMF can
affect the outcome in a systematic way. In addition, it is important
to point out that we can only give lower limits to the stellar mass
densities, as we are not able to detect heavily dust enshrouded
galaxies with large extinctions already at intermediate redshifts.

\begin{figure}
  \resizebox{\hsize}{!}{\includegraphics{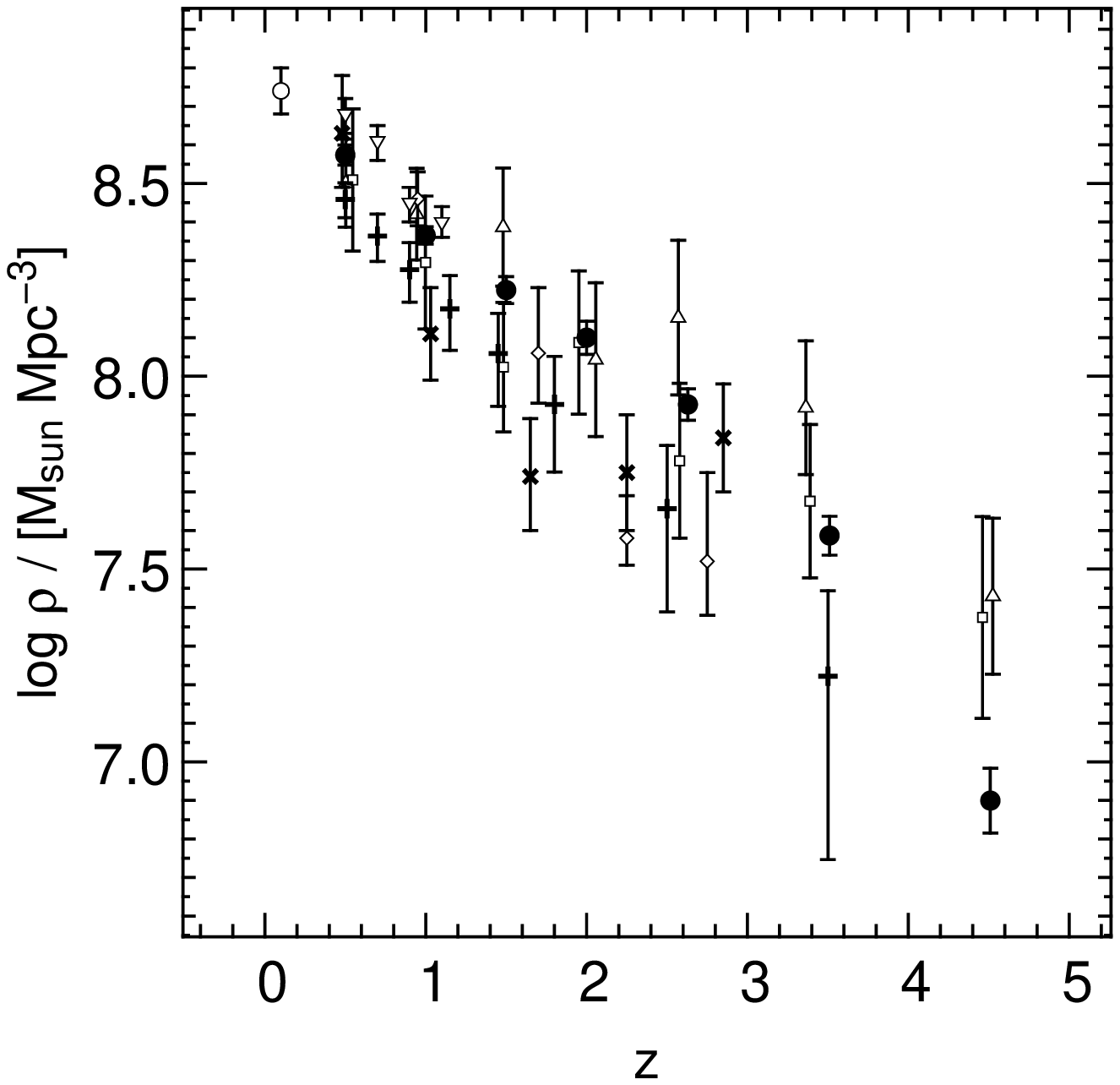}}
  \caption{Stellar mass densities as a function of redshift. We
  compare stellar mass densities derived in the present work on an
  area of about 140~$\mathrm{arcmin^2}$ utilizing mass limits from a
  $M/L$--ratio analysis (\emph{filled circles, see text}) to values
  from literature calculated without additional Spitzer/IRAC
  photometry. The results of \citealt{2005ApJ...619L.131D} (\emph{open
  triangles for FDF, 40~$\mathrm{arcmin^2}$; open squares for GSD,
  50~$\mathrm{arcmin^2}$}), \citealt{2003ApJ...587...25D} (\emph{open
  diamonds, from HDF--N, 5~$\mathrm{arcmin^2}$}),
  \citealt{2003ApJ...594L...9F} (\emph{tilted crosses, from HDF--S,
  5~$\mathrm{arcmin^2}$}), \citealt{2004ApJ...608..742D}
  (\emph{reversed triangles, from MUNICS survey,
  0.28~$\mathrm{deg^2}$}) and the local value \citep[][\emph{empty
  circle}]{2001MNRAS.326..255C} are shown. We also plot the
  datapoints of \citet{2006A&A...459..745F} (\emph{upright crosses,
  GOODS--S, 140~$\mathrm{arcmin^2}$}) as derived from the
  $\mathrm{K_S}$--selected subsample of the catalog including IRAC
  photometry by integrating the best fitting Schechter functions with
  smoothed parameters.}
  \label{fig:md}
\end{figure}

We list our results in Table~\ref{tab:md}, where we also compare the
mass densities to the local value derived by
\citet{2001MNRAS.326..255C}. It turns out that at a redshift of
$\langle z \rangle = 1$ at least 42~\% of today's stellar mass density
is already in place. This fraction decreases to 22~\% at $\langle z
\rangle = 2$ and about 6~\% at $\langle z \rangle = 3.5$. A comparison
with values from literature derived on the basis of a photometric
catalog without additional Spitzer/IRAC data is shown in
Fig.~\ref{fig:md}. Whilst up to intermediate redshifts the stellar
mass density is reproduced well (though with much smaller scatter
because of a larger observed area), we find systematic deviations at
high redshifts to lower densities, which one would expect from the
properties of the stellar mass functions discussed in Sect.~5.

A comparison of this work with the result of
\citet{2006A&A...459..745F}, who used the $\mathrm{K_S}$--selected
subsample of an almost identical catalog\footnote{Additional
spectroscopic redshift information for about 150 objects became
available in the meantime and were included in
\citet{2006A&A...459..745F}.} with about 3000 galaxies and integrated
the Schechter function using smoothed parameters, shows systematically
higher values for the stellar mass density. This trend is
strengthening from $\log \, \rho_\mathrm{this \;
work}/\rho_\mathrm{Fontana \; et \; al.} = 0. 11~\mathrm{dex}$ at a
redshift of $\langle z \rangle = 0.5$ to $0.31~\mathrm{dex}$ at
$\langle z \rangle = 3.5$. The discrepancy may have its origins in
slightly different model grids used to infer mass--to--light--ratios,
and in particular features of the utilized codes themselves.  In
contrast to the proceedings of \citet{2006A&A...459..745F}, we
restricted our template library to solar metallicity, but allowed for
an independent burst component when fitting the object luminosities
with model SEDs. As the difference in the inferred densities becomes
more pronounced with redshift, it stands to reason that the derived
masses of younger galaxies, in particular, are subject to systematic
deviations. In general, they are affected by larger errors as the
function of the mass--to--light--ratio steepens at low ages. Against
the background of the discussed degeneracies in age, metallicity and
extinction, differences in the underlying template models can affect
calculated stellar masses more distinctly here. Similarly, an
additional burst component can influence the derived stellar mass. If
a galaxy reveals both a high UV luminosity due to a recent starburst
and a large infrared luminosity i.e.\ substantial mass in an old
population, a single component fit may not be able to reproduce the
spectrum in the whole wavelength range simultaneously, as a young
model is too faint in the infrared and an old model not bright enough
in the blue. As a consequence, the inferred stellar mass can be lower
\citep{2007ApJ...655...51W}. Therefore, splitting up the fit in burst
and main component covers the range of mass--to--light--ratios in a
more flexible fashion.

\begin{table}
   \centering
   \caption{Stellar mass densities, as derived using the mass limits
   from a $M/L$--ratio analysis (\emph{upper section}) and a passively
   evolving scenario (\emph{lower section, see text}). The redshift
   interval $4.01 \le z < 5.01$ has not been corrected for
   completeness.}
   \label{tab:md}
   \begin{tabular}{cccc}
      \hline
      \hline
      redshift interval & $log \, \rho(z) / [M_{\sun} Mpc^{-3}]$ & $\sigma_{log \, \rho(z)}$ & $\frac{\rho(z)}{\rho(z = 0)}$\\
      \hline
      $0.25 \le z < 0.75$ & 8.57 & 0.03 & 68.2~\%\\
      $0.75 \le z < 1.25$ & 8.37 & 0.02 & 42.2~\%\\
      $1.25 \le z < 1.75$ & 8.22 & 0.03 & 30.4~\%\\
      $1.75 \le z < 2.25$ & 8.10 & 0.04 & 22.9~\%\\
      $2.25 \le z < 3.01$ & 7.93 & 0.04 & 15.4~\%\\
      $3.01 \le z < 4.01$ & 7.59 & 0.05 & 7.0~\%\\
      $4.01 \le z < 5.01$ & $>$ 6.90 & 0.08 & 1.4~\%\\
      \hline
      $0.25 \le z < 0.75$ & 8.57 & 0.03 & 68.1~\%\\
      $0.75 \le z < 1.25$ & 8.36 & 0.02 & 42.2~\%\\
      $1.25 \le z < 1.75$ & 8.22 & 0.03 & 30.4~\%\\
      $1.75 \le z < 2.25$ & 8.08 & 0.06 & 21.9~\%\\
      $2.25 \le z < 3.01$ & 7.89 & 0.06 & 14.1~\%\\
      $3.01 \le z < 4.01$ & 7.61 & 0.25 & 7.4~\%\\
      $4.01 \le z < 5.01$ & $>$ 6.90 & 0.08 & 1.4~\%\\
      \hline
   \end{tabular}
\end{table}

\section{Summary}

In this work we estimated the influence of newly available infrared
data longward of the K--filter on stellar mass estimates. To do so we
used the \emph{GOODS--MUSIC} catalog published by
\citet{2006A&A...449..951G}, which combines photometric data in 10
filters from 0.35 to 2.3~$\mu m$ with observations from the IRAC
instrument of the Spitzer space telescope at 3.6, 4.5, 5.8 and
8.0~$\mu m$. The catalog consists of 14\,847 objects within an area of
143.2~$\mathrm{arcmin^2}$ detected either in the z or the
$\mathrm{K_S}$--band. We computed stellar masses of this sample by
fitting stellar population synthesis models \citep{bruzual-2003-344}
to the data and multiplying the k--corrected absolute
$\mathrm{K_S}$--band luminosity with the $M/L_\mathrm{K}$--ratio of
the best fitting model SED. To probe the influence of the IRAC data on
the analysis we repeated the computation of stellar masses without
Spitzer photometry, keeping the photometric redshifts fixed, and
compared the outcome directly.

We found stellar masses to be overestimated on average, if further
constraining infrared data from Spitzer were not included in the
calculation. Whilst this trend is almost independent of mass itself, a
closer analysis reveals a strong dependency on redshift.  While up to
$z \approx 2$ the systematic deviation in mass is only moderate
($|\log \, M_\mathrm{U-4}/M_\mathrm{U-K}| < 0.2~\mathrm{dex}$) and
comparable to the intrinsic uncertainty of the method adopted to
estimate stellar masses, it increases strongly for higher redshifts
and reaches a maximum of a factor of three at $z \approx 3.5$. The
reason for this systematic shift can be traced back to insufficient
constraints on the slope of the spectra redward of the 4000~\AA\ break
at high redshifts. It turns out that, on average, models that are too
old, with excessively high absolute infrared luminosities and
$M/L$--ratios were assigned to the data if Spitzer photometry is not
included. Thus, a shift to lower stellar masses is likely to be
correlated with a decreasing age of the best fitting model SED. The
inclusion of one additional data--point longward of $3~\mu m$ can
already reduce the remaining error in mass significantly.

In the next step we used our results to calculate stellar mass
functions in different redshift intervals utilizing the
$V/V_\mathrm{max}$--formalism of \citet{1968ApJ...151..393S} to
correct our sample partly for incompleteness. To assign errors we
performed extensive Monte Carlo simulations where we considered
uncertainties in the underlying $M/L_\mathrm{K}$--ratios, redshifts
and a general error in photometry. Afterwards, the data points were
fitted via three free parameters using the analytical expression
suggested by \citet{1976ApJ...203..297S}. We found a pronounced
general decrease in numberdensity of all objects with redshift. Beyond
that, it is hard to say whether there is also a mass dependent
evolution. Although the change in computed Schechter parameter may
support this position, the effect can be caused by large--scale
structure as well.

Finally, we computed stellar mass densities as a function of
redshift. We summed the mass of our objects within each redshift
interval at the high mass end and integrated the Schechter function
derived on the basis of our stellar mass functions to complete the
result for lower masses. To estimate errors we again used Monte Carlo
simulations, but we pointed out that further effects such as biases in
stellar population synthesis models may be dominating sources of
uncertainty. By comparing the outcome to the local value of
\citet{2001MNRAS.326..255C} we found at least 42~\% of the stellar
mass density to be already in place at $\langle z \rangle = 1$. This
value decreases to 23~\% at $\langle z \rangle = 2$ and about 7~\% at
$\langle z \rangle = 3.5$. Therefore, up to intermediate redshifts our
results are in good agreement with values taken from literature
derived without additional Spitzer/IRAC data. However, at high
redshifts a systematic deviation to lower densities is present as one
would expect from the effect of Spitzer photometry on the calculation.

\begin{acknowledgements}
We thank the anonymous referee for the comments that helped to improve
the presentation of our results. We are grateful to Niv Drory for
providing the program used here to calculate stellar masses and for
valuable discussions at the final stage of this paper. We further
acknowledge \citet{2006A&A...449..951G} for making the
\emph{GOODS--MUSIC} catalog publicly available.
\end{acknowledgements}

\begin{appendix}
\section{Examples of SED--fits}

\begin{figure}
  \resizebox{\hsize}{!}{\includegraphics{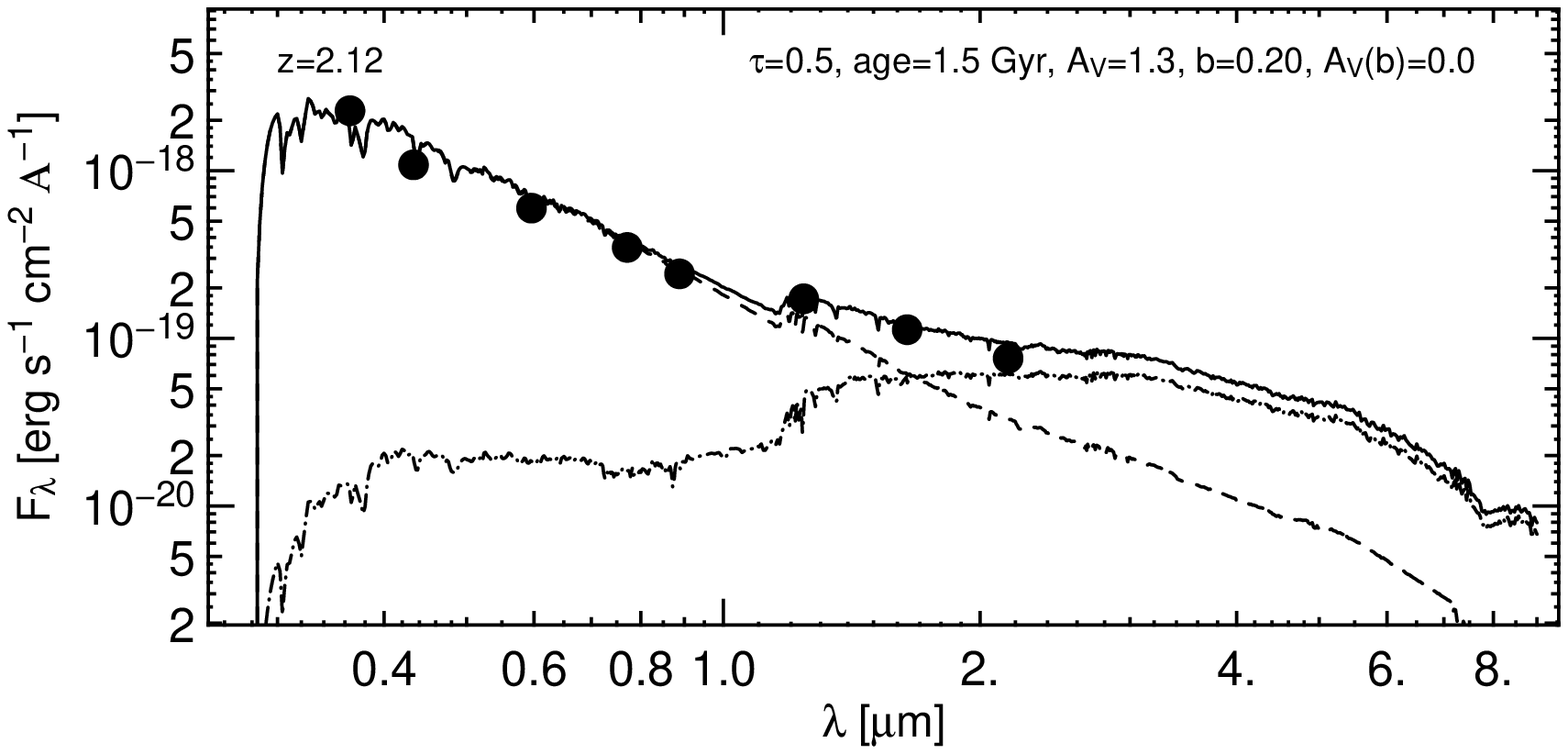}}
  \resizebox{\hsize}{!}{\includegraphics{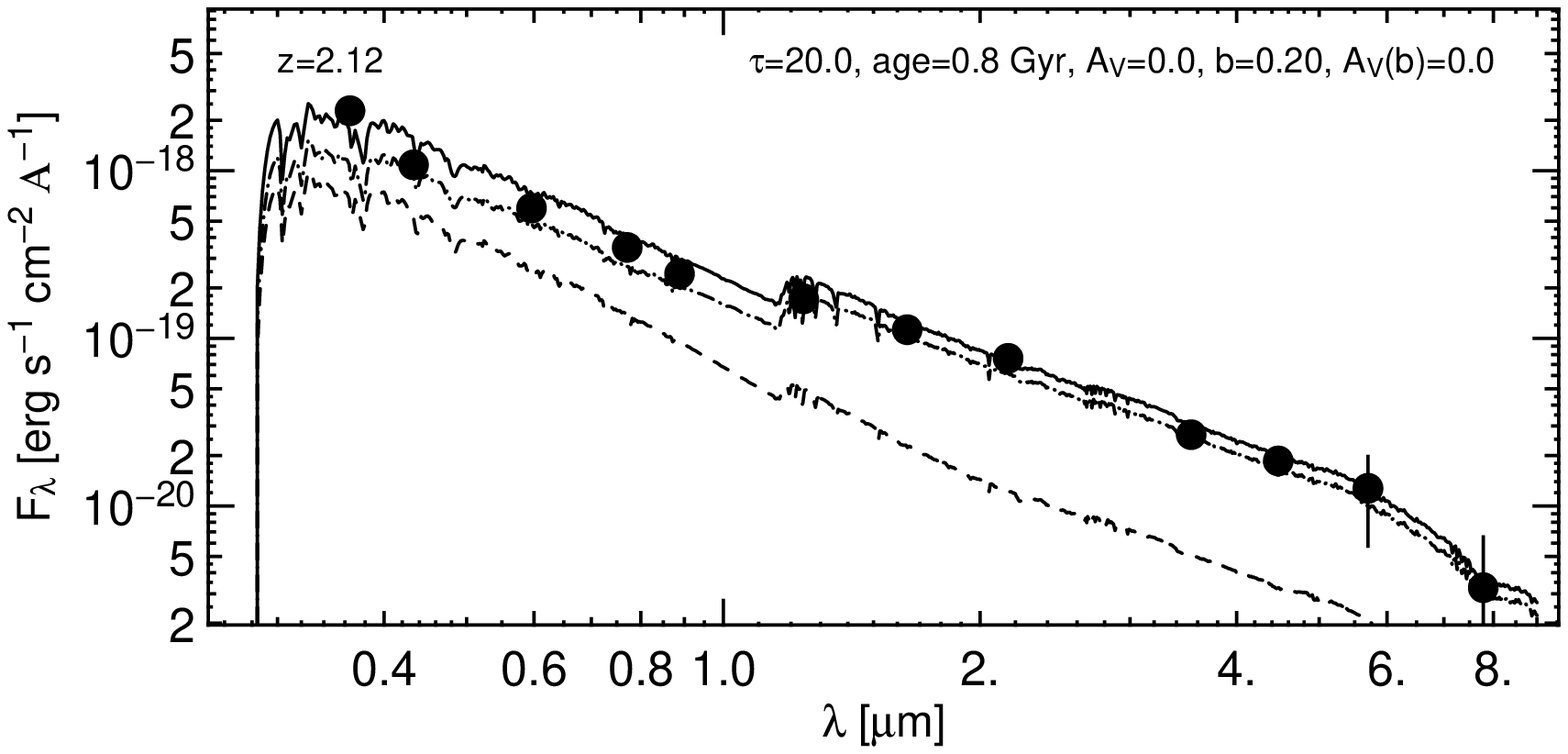}}
  \caption{Example of a change in the best fitting model SED
  (\emph{solid line}) due to inclusion of Spitzer photometry.
  \emph{Upper panel:} Without Spitzer/IRAC data the observed flux
  (\emph{filled circles}) is fitted with a young starburst
  (\emph{dashed line}) superimposed to a 1.5--Gyr--old main component
  with large extinction (\emph{dot--dashed line}). \emph{Lower panel:}
  After factoring in complementing infrared data, extinction decreases
  and the best fitting model SED becomes younger. The resulting
  stellar mass reduces from $\log \, M_\mathrm{U-K}/M_{\sun} = 10.32
  \; \mathrm{dex}$ to $\log \, M_\mathrm{U-4}/M_{\sun} = 9.69 \;
  \mathrm{dex}$.}
  \label{fig:ex_1}
\end{figure}

\begin{figure}
  \resizebox{\hsize}{!}{\includegraphics{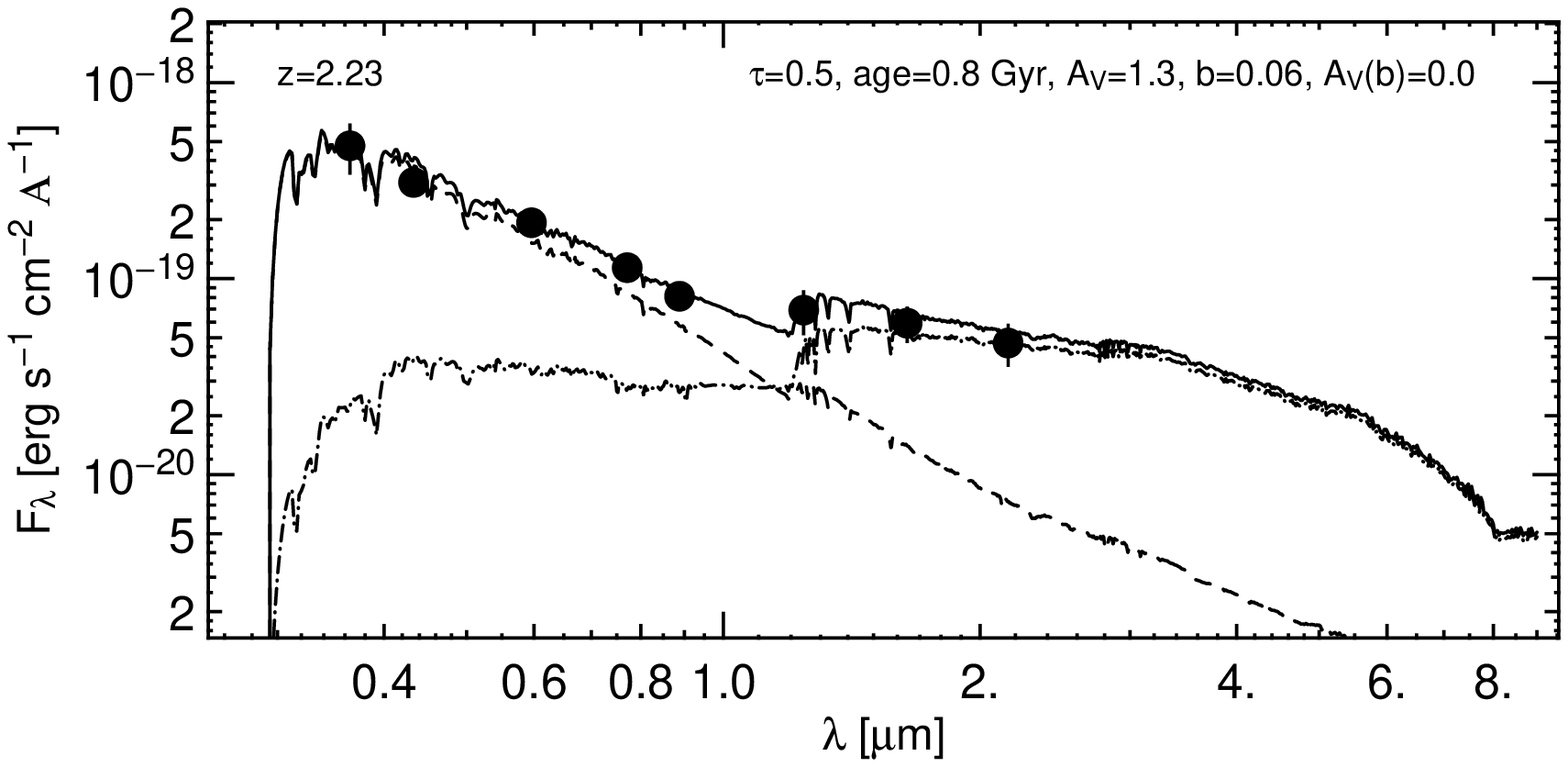}}
  \resizebox{\hsize}{!}{\includegraphics{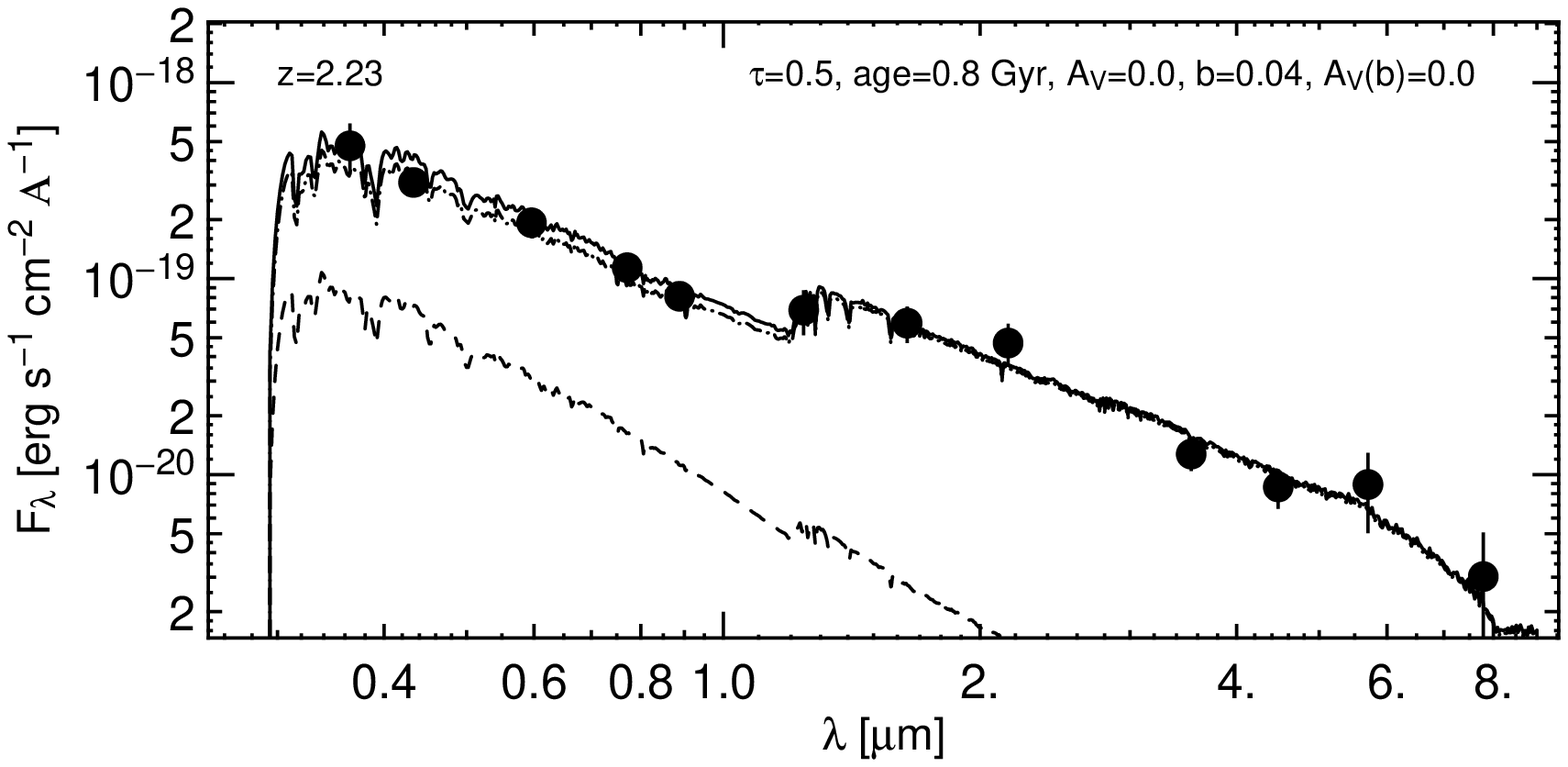}}
  \caption{Same as Fig.~\ref{fig:ex_1}. The change in the fit to
  photometric data where the extinction varies is shown. The resulting
  stellar mass reduces from $\log \, M_\mathrm{U-K}/M_{\sun} = 10.07
  \; \mathrm{dex}$ to $\log \, M_\mathrm{U-4}/M_{\sun} = 9.54 \;
  \mathrm{dex}$.}
  \label{fig:ex_2}
\end{figure}

\begin{figure}
  \resizebox{\hsize}{!}{\includegraphics{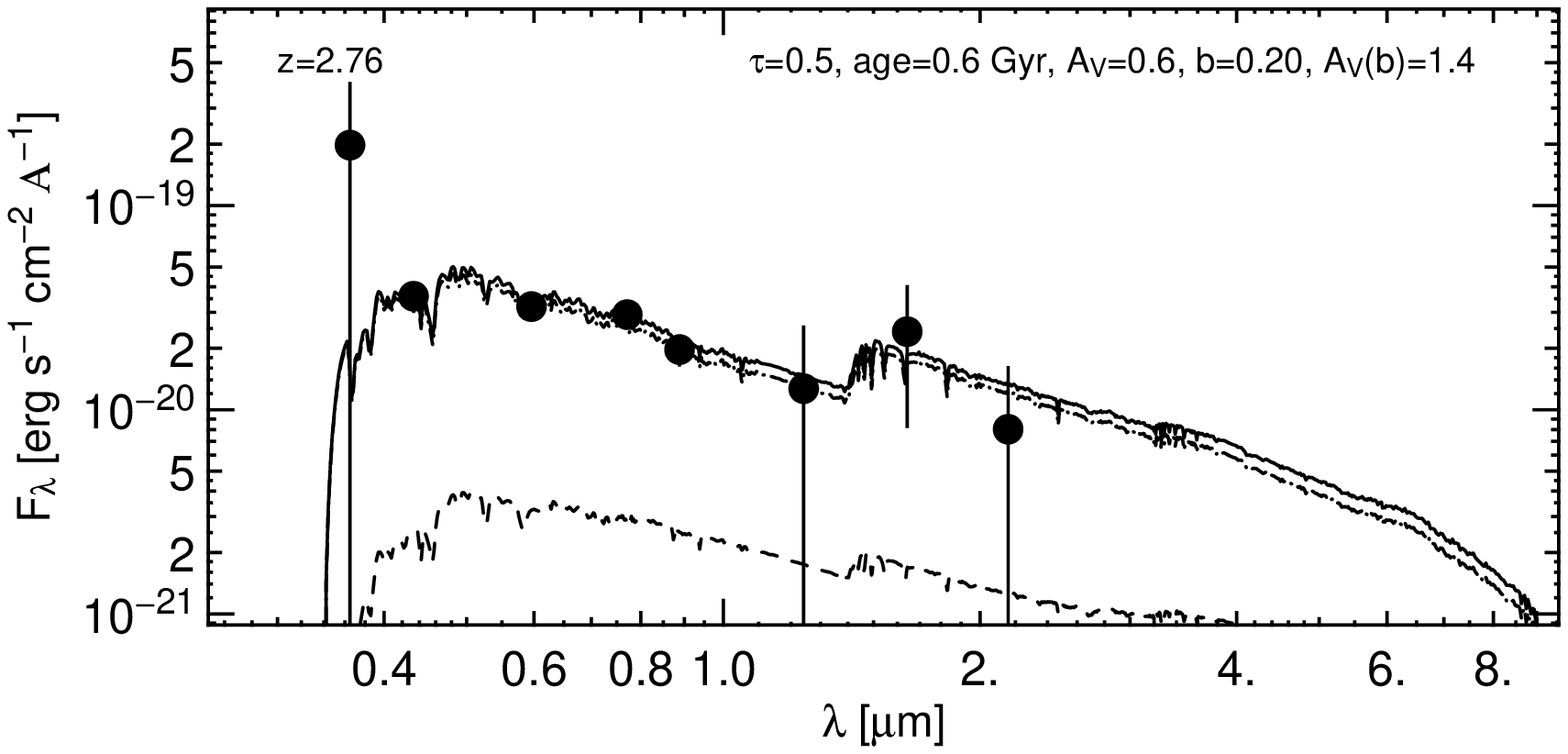}}
  \resizebox{\hsize}{!}{\includegraphics{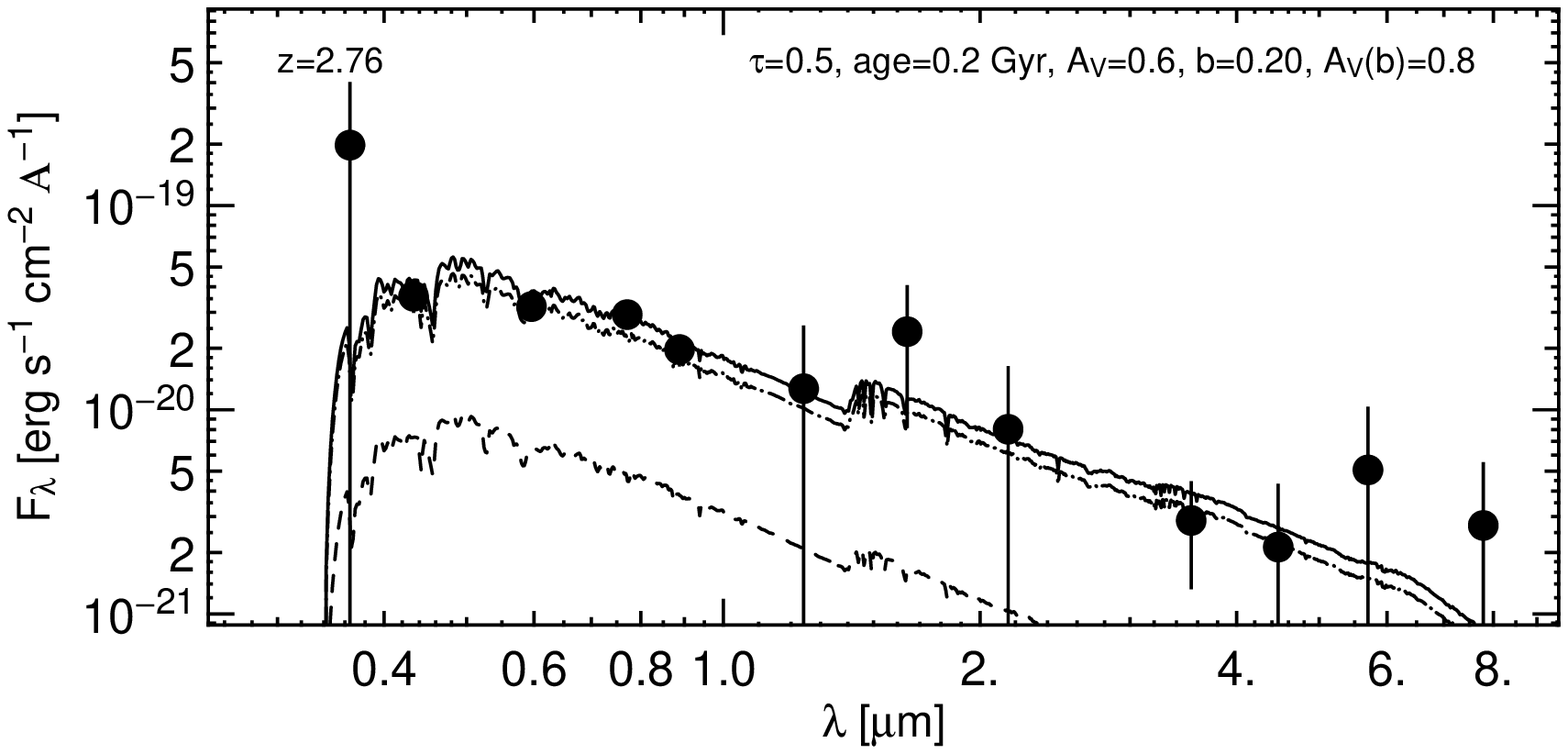}}
  \caption{Same as Fig.~\ref{fig:ex_1}. The change in the fit to
  photometric data where the absolute normalisation changes distinctly
  is shown. The resulting stellar mass decreases from $\log \,
  M_\mathrm{U-K}/M_{\sun} = 9.26 \; \mathrm{dex}$ to $\log \,
  M_\mathrm{U-4}/M_{\sun} = 8.52 \; \mathrm{dex}$.}
  \label{fig:ex_3}
\end{figure}

\begin{figure}
  \resizebox{\hsize}{!}{\includegraphics{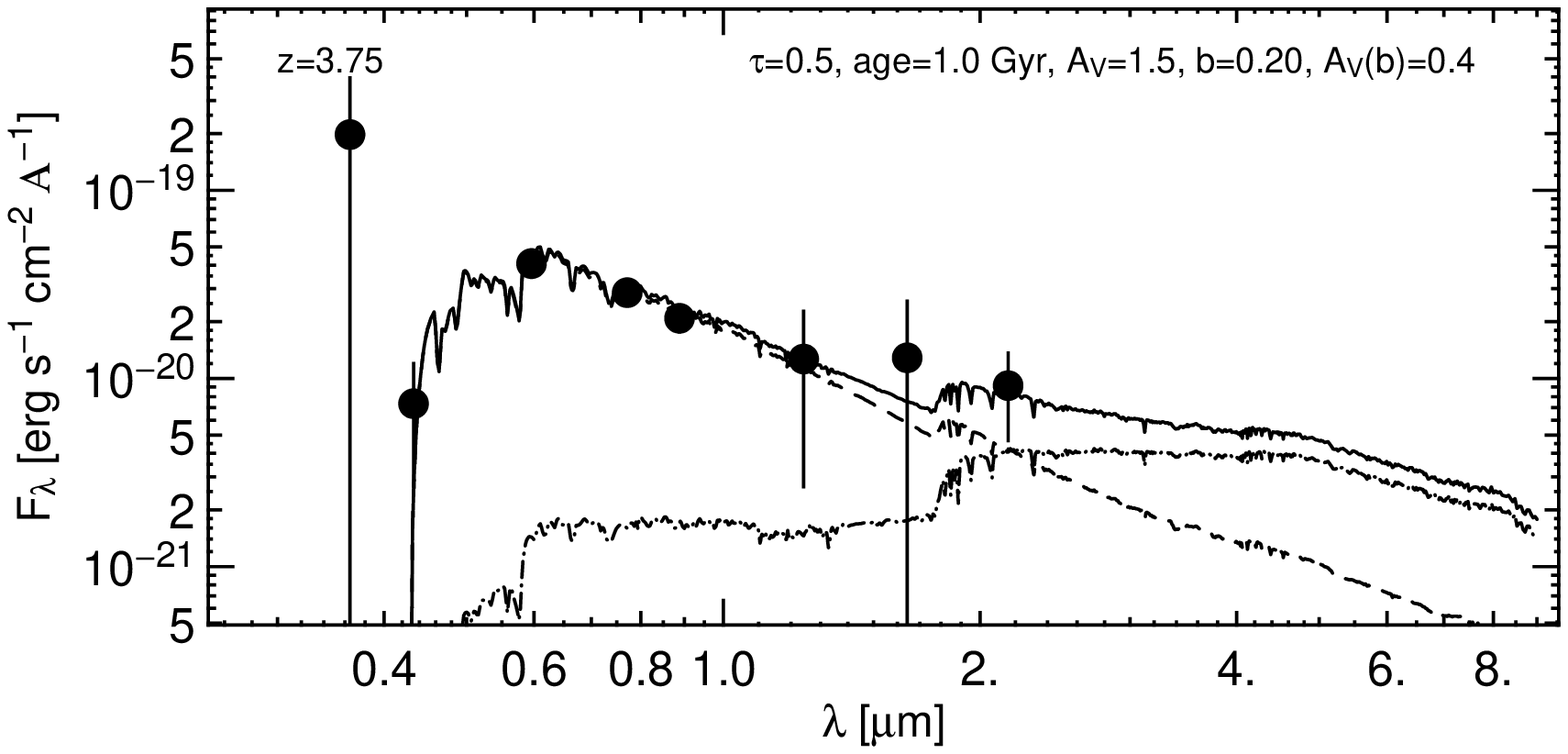}}
  \resizebox{\hsize}{!}{\includegraphics{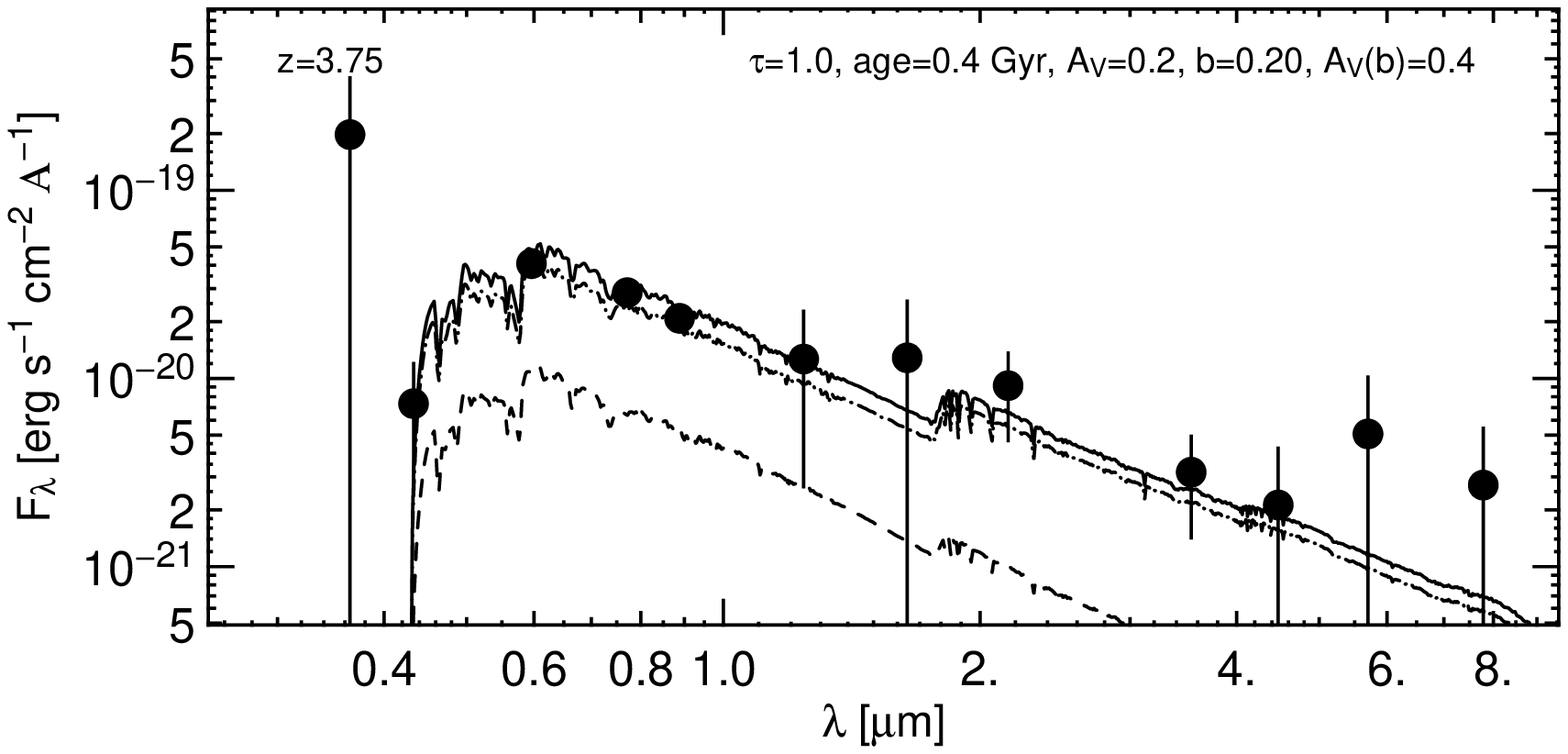}}
  \caption{Same as Fig.~\ref{fig:ex_1}. The figure shows an example
  of an insufficiently constrained optical-- to
  near--infrared--luminosity due to large photometric errors. The
  resulting stellar mass decreases from $\log \,
  M_\mathrm{U-K}/M_{\sun} = 9.45 \; \mathrm{dex}$ to $\log \,
  M_\mathrm{U-4}/M_{\sun} = 8.65 \; \mathrm{dex}$.}
  \label{fig:ex_4}
\end{figure}

We show some examples of a change in the best fitting model SED due to
inclusion of Spitzer photometry. The shift in calculated stellar
masses becomes manifest in a change of age, SFH and extinction
(Figs.~\ref{fig:ex_1}, \ref{fig:ex_2}) or the overall normalization
(Fig.~\ref{fig:ex_3}). In general, even a slight variation of the
$\mathrm{K_S}$--band luminosity can change the derived stellar mass
substantially if IRAC photometry is not taken into account. On the
other hand, we occasionally found the slope of the spectra at longer
wavelengths insufficiently constrained because of large photometric
errors (Fig.~\ref{fig:ex_4}).

\end{appendix}

\bibliographystyle{aa}
\bibliography{literature}

\begin{thebibliography}{28}
\expandafter\ifx\csname natexlab\endcsname\relax\def\natexlab#1{#1}\fi

\bibitem[{{Bell} {et~al.}(2003){Bell}, {McIntosh}, {Katz}, \&
  {Weinberg}}]{2003ApJS..149..289B}
{Bell}, E.~F., {McIntosh}, D.~H., {Katz}, N., \& {Weinberg}, M.~D. 2003, \apjs,
  149, 289

\bibitem[{{Borch} {et~al.}(2006){Borch}, {Meisenheimer}, {Bell}, {Rix}, {Wolf},
  {Dye}, {Kleinheinrich}, {Kovacs}, \& {Wisotzki}}]{2006A&A...453..869B}
{Borch}, A., {Meisenheimer}, K., {Bell}, E.~F., {et~al.} 2006, \aap, 453, 869

\bibitem[{{Brinchmann} \& {Ellis}(2000)}]{2000ApJ...536L..77B}
{Brinchmann}, J. \& {Ellis}, R.~S. 2000, \apjl, 536, L77

\bibitem[{{Bruzual}(2007)}]{2007astro.ph..2091B}
{Bruzual}, G. 2007, ArXiv Astrophysics e-prints

\bibitem[{{Bruzual} \& {Charlot}(2003)}]{bruzual-2003-344}
{Bruzual}, G. \& {Charlot}, S. 2003, \mnras, 344, 1000

\bibitem[{{Calzetti} {et~al.}(2000){Calzetti}, {Armus}, {Bohlin}, {Kinney},
  {Koornneef}, \& {Storchi-Bergmann}}]{2000ApJ...533..682C}
{Calzetti}, D., {Armus}, L., {Bohlin}, R.~C., {et~al.} 2000, \apj, 533, 682

\bibitem[{{Cole} {et~al.}(2001){Cole}, {Norberg}, {Baugh}, {Frenk},
  {Bland-Hawthorn}, {Bridges}, {Cannon}, {Colless}, {Collins}, {Couch},
  {Cross}, {Dalton}, {De Propris}, {Driver}, {Efstathiou}, {Ellis},
  {Glazebrook}, {Jackson}, {Lahav}, {Lewis}, {Lumsden}, {Maddox}, {Madgwick},
  {Peacock}, {Peterson}, {Sutherland}, \& {Taylor}}]{2001MNRAS.326..255C}
{Cole}, S., {Norberg}, P., {Baugh}, C.~M., {et~al.} 2001, \mnras, 326, 255

\bibitem[{{De Santis} {et~al.}(2006){De Santis}, {Grazian}, \&
  {Fontana}}]{2006MSAIS...9..454D}
{De Santis}, C., {Grazian}, A., \& {Fontana}, A. 2006, Memorie della Societa
  Astronomica Italiana Supplement, 9, 454

\bibitem[{{Dickinson} {et~al.}(2003{\natexlab{a}}){Dickinson}, {Giavalisco}, \&
  {The Goods Team}}]{dickinson}
{Dickinson}, M., {Giavalisco}, M., \& {The Goods Team}. 2003{\natexlab{a}}, in
  The Mass of Galaxies at Low and High Redshift, ed. R.~{Bender} \&
  A.~{Renzini}, 324--+

\bibitem[{{Dickinson} {et~al.}(2003{\natexlab{b}}){Dickinson}, {Papovich},
  {Ferguson}, \& {Budav{\'a}ri}}]{2003ApJ...587...25D}
{Dickinson}, M., {Papovich}, C., {Ferguson}, H.~C., \& {Budav{\'a}ri}, T.
  2003{\natexlab{b}}, \apj, 587, 25

\bibitem[{{Dopita} {et~al.}(2005){Dopita}, {Groves}, {Fischera}, {Sutherland},
  {Tuffs}, {Popescu}, {Kewley}, {Reuland}, \&
  {Leitherer}}]{2005ApJ...619..755D}
{Dopita}, M.~A., {Groves}, B.~A., {Fischera}, J., {et~al.} 2005, \apj, 619, 755

\bibitem[{{Drory} {et~al.}(2004{\natexlab{a}}){Drory}, {Bender}, {Feulner},
  {Hopp}, {Maraston}, {Snigula}, \& {Hill}}]{2004ApJ...608..742D}
{Drory}, N., {Bender}, R., {Feulner}, G., {et~al.} 2004{\natexlab{a}}, \apj,
  608, 742

\bibitem[{{Drory} {et~al.}(2004{\natexlab{b}}){Drory}, {Bender}, \&
  {Hopp}}]{2004ApJ...616L.103D}
{Drory}, N., {Bender}, R., \& {Hopp}, U. 2004{\natexlab{b}}, \apjl, 616, L103

\bibitem[{{Drory} {et~al.}(2001){Drory}, {Bender}, {Snigula}, {Feulner},
  {Hopp}, {Maraston}, {Hill}, \& {Mendes de Oliveira}}]{2001ApJ...562L.111D}
{Drory}, N., {Bender}, R., {Snigula}, J., {et~al.} 2001, \apjl, 562, L111

\bibitem[{{Drory} {et~al.}(2005){Drory}, {Salvato}, {Gabasch}, {Bender},
  {Hopp}, {Feulner}, \& {Pannella}}]{2005ApJ...619L.131D}
{Drory}, N., {Salvato}, M., {Gabasch}, A., {et~al.} 2005, \apjl, 619, L131

\bibitem[{{Dunlop} {et~al.}(2007){Dunlop}, {Cirasuolo}, \&
  {McLure}}]{2006astro.ph..6192D}
{Dunlop}, J.~S., {Cirasuolo}, M., \& {McLure}, R.~J. 2007, \mnras, 376, 1054

\bibitem[{{Fontana} {et~al.}(2003){Fontana}, {Donnarumma}, {Vanzella},
  {Giallongo}, {Menci}, {Nonino}, {Saracco}, {Cristiani}, {D'Odorico}, \&
  {Poli}}]{2003ApJ...594L...9F}
{Fontana}, A., {Donnarumma}, I., {Vanzella}, E., {et~al.} 2003, \apjl, 594, L9

\bibitem[{{Fontana} {et~al.}(2004){Fontana}, {Pozzetti}, {Donnarumma},
  {Renzini}, {Cimatti}, {Zamorani}, {Menci}, {Daddi}, {Giallongo}, {Mignoli},
  {Perna}, {Salimbeni}, {Saracco}, {Broadhurst}, {Cristiani}, {D'Odorico}, \&
  {Gilmozzi}}]{2004A&A...424...23F}
{Fontana}, A., {Pozzetti}, L., {Donnarumma}, I., {et~al.} 2004, \aap, 424, 23

\bibitem[{{Fontana} {et~al.}(2006){Fontana}, {Salimbeni}, {Grazian},
  {Giallongo}, {Pentericci}, {Nonino}, {Fontanot}, {Menci}, {Monaco},
  {Cristiani}, {Vanzella}, {de Santis}, \& {Gallozzi}}]{2006A&A...459..745F}
{Fontana}, A., {Salimbeni}, S., {Grazian}, A., {et~al.} 2006, \aap, 459, 745

\bibitem[{{Grazian} {et~al.}(2006){Grazian}, {Fontana}, {de Santis}, {Nonino},
  {Salimbeni}, {Giallongo}, {Cristiani}, {Gallozzi}, \&
  {Vanzella}}]{2006A&A...449..951G}
{Grazian}, A., {Fontana}, A., {de Santis}, C., {et~al.} 2006, \aap, 449, 951

\bibitem[{{Maraston} {et~al.}(2006){Maraston}, {Daddi}, {Renzini}, {Cimatti},
  {Dickinson}, {Papovich}, {Pasquali}, \& {Pirzkal}}]{2006ApJ...652...85M}
{Maraston}, C., {Daddi}, E., {Renzini}, A., {et~al.} 2006, \apj, 652, 85

\bibitem[{{Rudnick} {et~al.}(2003){Rudnick}, {Rix}, {Franx}, {Labb{\'e}},
  {Blanton}, {Daddi}, {F{\"o}rster Schreiber}, {Moorwood}, {R{\"o}ttgering},
  {Trujillo}, {van de Wel}, {van der Werf}, {van Dokkum}, \& {van
  Starkenburg}}]{2003ApJ...599..847R}
{Rudnick}, G., {Rix}, H.-W., {Franx}, M., {et~al.} 2003, \apj, 599, 847

\bibitem[{{Schechter}(1976)}]{1976ApJ...203..297S}
{Schechter}, P. 1976, \apj, 203, 297

\bibitem[{{Schmidt}(1968)}]{1968ApJ...151..393S}
{Schmidt}, M. 1968, \apj, 151, 393

\bibitem[{{Somerville} {et~al.}(2004){Somerville}, {Lee}, {Ferguson},
  {Gardner}, {Moustakas}, \& {Giavalisco}}]{2004ApJ...600..171S}
{Somerville}, R.~S., {Lee}, K., {Ferguson}, H.~C., {et~al.} 2004, \apj, 600,
  171

\bibitem[{{van der Wel} {et~al.}(2006){van der Wel}, {Franx}, {Wuyts}, {van
  Dokkum}, {Huang}, {Rix}, \& {Illingworth}}]{2006ApJ...652...97V}
{van der Wel}, A., {Franx}, M., {Wuyts}, S., {et~al.} 2006, \apj, 652, 97

\bibitem[{{Werner} {et~al.}(2004){Werner}, {Roellig}, {Low}, {Rieke}, {Rieke},
  {Hoffmann}, {Young}, {Houck}, {Brandl}, {Fazio}, {Hora}, {Gehrz}, {Helou},
  {Soifer}, {Stauffer}, {Keene}, {Eisenhardt}, {Gallagher}, {Gautier}, {Irace},
  {Lawrence}, {Simmons}, {Van Cleve}, {Jura}, {Wright}, \&
  {Cruikshank}}]{2004ApJS..154....1W}
{Werner}, M.~W., {Roellig}, T.~L., {Low}, F.~J., {et~al.} 2004, \apjs, 154, 1

\bibitem[{{Wuyts} {et~al.}(2007){Wuyts}, {Labb{\'e}}, {Franx}, {Rudnick}, {van
  Dokkum}, {Fazio}, {F{\"o}rster Schreiber}, {Huang}, {Moorwood}, {Rix},
  {R{\"o}ttgering}, \& {van der Werf}}]{2007ApJ...655...51W}
{Wuyts}, S., {Labb{\'e}}, I., {Franx}, M., {et~al.} 2007, \apj, 655, 51

\end{thebibliography}

\end{document}